\documentclass[prd, reprint, amsmath, amssymb, nofootinbib, floatfix, aps, longbibliography, superscriptaddress]{revtex4-1}
\usepackage{graphicx}
\usepackage{dcolumn}
\usepackage{bm}
\usepackage[colorlinks=true,urlcolor=rossos,anchorcolor=black,citecolor=mygreen,filecolor=black,linkcolor=black,menucolor=blue,linktocpage=true]{hyperref} 

\usepackage{subfigure}
\usepackage[utf8]{inputenc} 

\usepackage{booktabs}
\usepackage{colortbl}
\usepackage{collcell}
\usepackage{enumerate}
\usepackage{float}
\usepackage{verbatim}
\usepackage{multirow}
\usepackage{xparse}
\usepackage{tabularx}

\graphicspath{{figs/}}

\usepackage{multirow}

\definecolor{rossos}{cmyk}{0,1,1,0.55}
\definecolor{mygreen}{rgb}{0.27, 0.64, 0.48}
\definecolor{mygray}{gray}{.95}

\definecolor{revred}{rgb}{0.85,0,0}
\newcommand{\rev}[1]{{\color{revred}\ifmmode #1\else \bfseries #1\fi}}

\begin{document}

\title{Circular polarization effects induced by photon-axion mixing in astrophysical environments}


\author{Meng Wang}
\affiliation{School of Physical Science and Technology, Kunming University, Kunming 650214, China}

\author{Nan Ding}
\email{Corresponding author: orient.dn@foxmail.com}
\affiliation{School of Physical Science and Technology, Kunming University, Kunming 650214, China}

\author{Qiusheng Gu}
\affiliation{School of Astronomy and Space Science, Nanjing University, Nanjing, Jiangsu 210093, China}

\author{Yunyong Tang}
\affiliation{School of Physical Science and Technology, Kunming University, Kunming 650214, China}

\author{Rui Xue}
\affiliation{Department of Physics, Zhejiang Normal University, Jinhua 321004, China}

\date{\today}

\begin{abstract}

Axions and axion-like particles (ALPs) are compelling candidates for dark matter and new physics beyond the Standard Model. Photon-axion mixing in external magnetic fields not only modifies the photon energy spectrum and linear polarization state but also induces circular polarization signals. Compared to spectral and linear polarization methods, circular polarization benefits from lower astrophysical background contamination and weaker dependence on the intrinsic source spectrum, providing an independent probe for axion searches. In this work, we study the circular polarization induced by photon-axion mixing within the chiral basis framework. By analytically solving the evolution equations under the single-domain approximation, we derive an expression for the circular polarization degree $P_C$, applicable in the resonant, strong coupling, and weak coupling regimes. The opposite-phase coupling of the axion field to left- and right-handed circular polarization components generates phase differences and intensity asymmetries, thereby converting initially linearly polarized light into non-zero circular polarization signals. Within single-domain magnetic field models, we compare the energy-dependent circular polarization in four astrophysical environments (active galactic nucleus jets, the intracluster medium, the intergalactic medium, and the Galactic magnetic fields). We find that the X-ray to MeV band represents the most sensitive observational window for axion-induced circular polarization signals. In multi-domain propagation models, using the blazar S4 0954+65 as a case study, phase accumulation in random magnetic domains causes the circular polarization degree to fluctuate with redshift and exhibit pronounced energy structures in the X-ray to MeV band. Using the optical circular polarization upper limit $P_C < 0.184\%$ (measured in the z-SDSS band) from this source, we statistically constrain $g_{a\gamma\gamma} \lesssim 3\times 10^{-11}\,\mathrm{GeV}^{-1}$ (95\% confidence level) for $m_a \sim 10^{-16}$--$10^{-10}\ \mathrm{eV}$, with the strongest constraint reaching $g_{a\gamma\gamma}\sim6\times10^{-12}\,\mathrm{GeV}^{-1}$ near $m_a \sim 10^{-14}\,\mathrm{eV}$. These results establish circular polarization as a complementary axion probe, and future high-energy circular polarization observations are expected to further strengthen constraints on the ultralight axion parameter space.

\end{abstract}

\maketitle

\section{Introduction}
The particle nature of dark matter remains one of the central open questions in modern cosmology and particle physics. Among numerous candidate particles, axions and axion-like particles (ALPs) have attracted significant attention due to their compelling theoretical motivation. The QCD axion was originally introduced via the Peccei-Quinn mechanism to address the strong CP problem in quantum chromodynamics (QCD)~\cite{Peccei:1977hh}. More generally, theoretical frameworks beyond the Standard Model, such as string theory, predict a large number of light pseudoscalar particles, namely ALPs~\cite{Svrcek:2006yi}. Unlike the QCD axion, the mass and coupling constant of ALPs can be independent of each other, thereby covering a much broader parameter space and yielding diverse signatures across dark matter searches, cosmological evolution, and high-energy astrophysics ~\cite{Cadamuro:2012rm}. Unless otherwise specified, we use axions throughout this paper to collectively refer to both the QCD axion and ALPs.

The feeble coupling between axions and electromagnetic fields enables their mutual conversion with photons in external magnetic fields, thereby modifying the photon flux, energy spectrum, and polarization state~\cite{maiani1986effects,PhysRevD.37.1237,Roncadelli:2012ar,Bassan:2010ya,Finelli:2008jv}. This mechanism underpins laboratory searches including the axion dark matter experiment (ADMX)~\cite{du2018search}, the Any Light Particle Search II (ALPS-II)~\cite{bahre2013any}, and the CERN Axion Solar Telescope  (CAST)~\cite{anastassopoulos2017new}, and motivates searches in astrophysical environments with strong magnetic fields and long propagation baselines. Currently, astrophysical searches primarily rely on two classes of observables: spectral structure and polarization information. Spectral methods search for energy-dependent oscillations, spectral irregularities, or flux anomalies in the X-ray, gamma-ray, or radio bands, and have been applied to the Galactic Center, active galactic nuclei, super star clusters, and pulsars~\cite{Zhu:2024kmu,Reynes:2021bpe,Reynolds:2019uqt,Dekker:2025vcg,Dessert:2020lil,Noordhuis:2022ljw}. However, spectral constraints depend on assumptions about the intrinsic source spectrum, absorption backgrounds, and magnetic field configurations, which introduces unavoidable systematic uncertainties. 

Polarization observations, which encode the coherent phase of the electromagnetic field and the anisotropic response of the propagation medium, offer complementary probes of axion-photon mixing through two distinct mechanisms. First, photon-axion conversion in an external magnetic field selectively affects the photon polarization component parallel to the transverse magnetic field, thereby inducing or modifying linear polarization. Magnetic white dwarfs, neutron-star magnetospheres, and gamma-ray bursts have been used in this context to constrain axion-photon couplings over different mass ranges~\cite{Dessert:2022yqq,Benabou:2025jcv,Song:2024rru,betancourt2025polarization}. Second, if ultralight axions constitute a coherent dark matter background, the oscillating axion field can induce time-dependent birefringence and periodic variations of the polarization angle~\cite{Liu:2019brz}. Based on this mechanism, pulsar polarimetry arrays, long-term optical polarization monitoring of blazars, and millimeter-wave polarization calibration sources have all been employed to search for polarization angle oscillations induced by axion dark matter~\cite{liu2023pulsar,PPTA:2024mgh,li2026probing,Wang:2024sdz,Huang:2025rrb,POLARBEAR:2025djl}. Nevertheless, linear polarization and polarization-angle observables are often degenerate with magnetic-field geometry in the emission region, intrinsic synchrotron polarization, Faraday rotation, and source variability. Therefore, it remains necessary to develop observables with lower backgrounds and complementary to linear polarization.

In many standard astrophysical emission processes, the degree of circular polarization is typically much lower than that of linear polarization. Circular polarization can therefore partly avoid the above degeneracies and provides a new channel for probing axions. Several recent works have examined axion-induced circular polarization from specific angles. Masaki et al.~\cite{Masaki:2017aea} showed how magnetic field geometry shapes polarization dynamics and the associated observational constraints. Yao et al.~\cite{Yao:2022col} derived analytical expressions for blazar circular polarization in the weak-mixing limit and constrained the photon-axion coupling from optical observations; a follow-up study demonstrated that coherent ALP dark matter oscillations can resonantly enhance this conversion via a Floquet mechanism~\cite{Yao:2026yez}.  Separately, propagation through stochastic magnetic fields was shown to generate nontrivial circular polarization even for initially unpolarized photon beams~\cite{Chiba:2025hgr}. Collectively, these studies demonstrate that circular polarization serves not only as a direct propagation effect of axion-photon mixing, but also as a novel observable complementary to energy spectra, linear polarization, and time series of polarization angles.

Despite these advances, several aspects remain to be addressed. Existing analytical treatments are largely developed in the linear polarization basis or within the weak-mixing approximation, which obscures the chiral phase structure of the left- and right-handed circular polarization components. Furthermore, most studies are restricted to single-domain or idealized multi-domain magnetic field configurations, leaving the cumulative evolution of circular polarization along realistic multi-environment lines of sight systematically unexplored. Observational constraints have also been obtained predominantly in the optical band, whereas the photon-axion mixing efficiency is strongly energy-dependent and is expected to peak in the X-ray to MeV regime. 

Motivated by these considerations, in this work we investigate circular polarization induced by axion-photon mixing in astrophysical environments. We formulate the mixing problem in the chiral basis and derive analytical expressions for the circular polarization degree in the weak-mixing, strong-mixing, and resonant regimes, making explicit how the opposite phase responses of the left- and right-handed circular polarization modes generate circular polarization during propagation. We then examine the energy dependence of the signal in representative single-domain environments, including AGN jets, the intracluster medium, the intergalactic medium, and the Galactic magnetic field. To connect the analytical results with realistic propagation, we further construct a multi-domain model along the line of sight toward the blazar S4 0954+65 and use the optical upper limit on its circular polarization to constrain the axion parameter space. 

The paper is organized as follows. In Sec.~\ref{sec:Analytic model},  we develop the chiral-basis formalism and derive the analytical expressions for the circular polarization degree. In Sec.~\ref{sec:three}, we apply the single-domain model to several representative astrophysical environments and examine the corresponding energy dependence. In Sec.~\ref{sec:four}, we study multi-domain propagation toward S4 0954+65 and derive constraints from the optical circular-polarization limit. We summarize our results in Sec.~\ref{sec:five}.

\section{Analytic model for circular polarization in photon-axion mixing}
\label{sec:Analytic model}
In this section, we present an analytic model for circular polarization in a photon-axion mixing system, formulated in a chiral basis. Solving the evolution equations for a single-domain magnetic field, we derive closed-form expression for the circular polarization degree \(P_C(z)\) and discuss its simplified forms under several representative physical conditions.

The photon-axion coupling is described by the Lagrangian
\begin{equation}
\begin{split}
\mathcal{L}_{\text{ALP}} &= \frac{1}{2} \partial^\mu a \partial_\mu a - \frac{1}{2} m_a^2 a^2 - \frac{1}{4} g_{a\gamma\gamma} a F_{\mu\nu} \tilde{F}^{\mu\nu} \\
&= \frac{1}{2} \partial^\mu a \partial_\mu a - \frac{1}{2} m_a^2 a^2 + g_{a\gamma\gamma} a \mathbf{E} \cdot \mathbf{B},
\end{split}
\end{equation}
where \(a\) is the axion field with mass \(m_a\),  \(g_{a\gamma\gamma}\) is the photon-axion coupling constant,  \(F_{\mu\nu}\) is the electromagnetic tensor, and \(\tilde{F}^{\mu\nu}\) is its dual. We consider a photon-axion state with energy \(E\) propagating along the \(z\)-axis through a magnetized medium. In the short-wavelength approximation, the evolution of the system is governed by a Schrödinger-like equation~\cite{PhysRevD.37.1237}
\begin{equation}
i\frac{d}{dz}\psi(z) = \mathcal{H}(E,z)\,\psi(z).
\label{eq:evolution}
\end{equation}

Photon-axion mixing occurs in the transverse magnetic field \(\mathbf{B}_T\) perpendicular to the propagation direction.  Taking \(\mathbf{B}_T\) to be oriented along the \(y\)-axis, the mixing matrix in the linear polarization basis \(\psi = (|x\rangle, |y\rangle, |a\rangle)^{\mathrm{T}}\) reads~\cite{Bassan:2010ya}
\begin{equation}
\mathcal{H}_{\text{lin}} = \begin{pmatrix}
\Delta_{\perp} & 0 & 0 \\
0 & \Delta_{\parallel} & \Delta_{a\gamma} \\
0 & \Delta_{a\gamma} & \Delta_a
\end{pmatrix},
\end{equation}
where the matrix elements include plasma effects, the QED vacuum polarization, the photon-axion interaction, and the axion mass effect.
To investigate the circular polarization evolution of photons, we transform the system into the chiral basis defined by 
\begin{equation}
|R\rangle = \frac{1}{\sqrt{2}}\bigl(|x\rangle + i|y\rangle\bigr),\;\;
|L\rangle = \frac{1}{\sqrt{2}}\bigl(|x\rangle - i|y\rangle\bigr).
\end{equation}
The corresponding unitary transformation matrix is
\begin{equation}
U = \frac{1}{\sqrt{2}}
\begin{pmatrix}
1 & i & 0 \\
1 & -i & 0 \\
0 & 0 & \sqrt{2}
\end{pmatrix}.
\end{equation}
In the chiral basis $\psi = (|R\rangle,\,|L\rangle,\,|a\rangle)^{\mathrm{T}}$, 
the mixing matrix is obtained via the unitary transformation 
\begin{equation}
\psi_{\text{chiral}} = U \psi_{\text{lin}},\;\;
H_{\text{chiral}} = U H_{\text{lin}} U^\dagger.
\end{equation}
With this convention, one obtains
\begin{equation}
\mathcal{H}_{\text{chiral}} = 
\begin{pmatrix}
\frac{\Delta_\perp + \Delta_\parallel}{2} & \frac{\Delta_\perp - \Delta_\parallel}{2} & \frac{i\Delta_{a\gamma}}{\sqrt{2}} \\
\frac{\Delta_\perp - \Delta_\parallel}{2} & \frac{\Delta_\perp + \Delta_\parallel}{2} & -\frac{i\Delta_{a\gamma}}{\sqrt{2}} \\
-\frac{i\Delta_{a\gamma}}{\sqrt{2}} & \frac{i\Delta_{a\gamma}}{\sqrt{2}} & \Delta_a
\end{pmatrix},
\end{equation}
where the photon-axion coupling term is $\Delta_{a\gamma} \equiv g_{a\gamma\gamma}B_T/2$ and the axion mass term is $\Delta_a \equiv -m_a^2/(2E)$. The photon-medium interaction terms 
$\Delta_{\perp}$ and $\Delta_{\parallel}$ are defined as~\cite{Galanti:2024lfn}
\begin{equation}
\Delta_{\perp} \equiv -\frac{\omega_{\mathrm{pl}}^2}{2E} 
+ \frac{2\alpha}{45\pi}\left(\frac{B_T}{B_{\mathrm{cr}}}\right)^2 E,
\end{equation}
\begin{equation}
\Delta_{\parallel} \equiv -\frac{\omega_{\mathrm{pl}}^2}{2E} 
+ \frac{7\alpha}{90\pi}\left(\frac{B_T}{B_{\mathrm{cr}}}\right)^2 E, 
\end{equation}
where $\alpha$ is the fine-structure constant, $\omega_{\mathrm{pl}} = \sqrt{4\pi\alpha n_e/m_e}$ is the plasma frequency, 
and $B_{\mathrm{cr}} = 4.41\times10^{13}\,\mathrm{G}$ is the critical magnetic field.

For the astrophysical scenarios considered here (magnetic fields from  $\mu\mathrm{G}$ to $\mathrm{G}$,  photon energies  in the keV-MeV range), the QED vacuum polarization term is negligible compared with the plasma term 
$\Delta_{\mathrm{pl}} = -\omega_{\mathrm{pl}}^2/(2E)$. Consequently, one may approximate \( \Delta_{\perp} = \Delta_{\parallel} = \Delta_{pl} \), and the mixing matrix can be rewritten as
\begin{equation}
\mathcal{H}_{\mathrm{chiral}} =
\begin{pmatrix}
\Delta_{\mathrm{pl}} & 0 & \dfrac{i\Delta_{a\gamma}}{\sqrt{2}} \\[8pt]
0 & \Delta_{\mathrm{pl}} & -\dfrac{i\Delta_{a\gamma}}{\sqrt{2}} \\[8pt]
-\dfrac{i\Delta_{a\gamma}}{\sqrt{2}} & \dfrac{i\Delta_{a\gamma}}{\sqrt{2}} & \Delta_a
\end{pmatrix}.
\label{9}
\end{equation}
This form shows explicitly that left- and right-circularly polarization states couple 
to the axion with opposite phases. Such a chirality-dependent coupling can convert an initially linearly polarized photon beam into a state with nonzero circular polarization, 
providing an observable handle for axion detection.

The evolution equation Eq.~\eqref{eq:evolution} is solved by diagonalizing the mixing matrix 
(see appendix \ref{A} for more details). Considering a uniform magnetic domain with no initial axion component, the initial photon state is taken to be 
\begin{equation}
|\psi(0)\rangle = C_R|R\rangle + C_L|L\rangle, \;\; C_a = 0.
\end{equation}
The degree of circular polarization after the photon has propagated a distance $z$ 
is defined by
\begin{equation}
P_C(z) = \frac{|R(z)|^2 - |L(z)|^2}{|R(z)|^2 + |L(z)|^2}, 
\;\; P_C \in [-1,\,1].
\label{11}
\end{equation}

For an initially linearly polarized photon state with polarization angle $\beta$ relative to the $x$-axis, the circular polarization degree $P_C(z)$ can be expressed as
\begin{equation}
P_C(z) = \frac{2\sin(2\beta)\bigl(A\sin\phi_{12} + A'\sin\phi_{13}\bigr)}{\cos^2\!\beta + 4\sin^2\!\beta\,(A^2 + A'^2) + 8AA'\sin^2\!\beta\,\cos\phi_{23}},
\label{12}
\end{equation}
with
\[
A = \frac{2\Delta_{a\gamma}^2}{\bigl(\Delta_a - \Delta_{\mathrm{pl}} - \Delta_{\mathrm{osc}}\bigr)^2 + 4\Delta_{a\gamma}^2},\]
\[
A' = \frac{2\Delta_{a\gamma}^2}{\bigl(\Delta_a - \Delta_{\mathrm{pl}} + \Delta_{\mathrm{osc}}\bigr)^2 + 4\Delta_{a\gamma}^2}.
\]
The phase differences $\phi_{ij}$ arising from interference between distinct eigenmodes read
\[
\phi_{12} = \frac{1}{2}\bigl(\Delta_a - \Delta_{\mathrm{pl}} - \Delta_{\mathrm{osc}}\bigr)z,\]
\[
\phi_{13} = \frac{1}{2}\bigl(\Delta_a - \Delta_{\mathrm{pl}} + \Delta_{\mathrm{osc}}\bigr)z,\]
\[
\phi_{23} = \Delta_{\mathrm{osc}}\,z,\]
with oscillation frequency
\[
\Delta_{\mathrm{osc}} = \sqrt{\bigl(\Delta_a - \Delta_{\mathrm{pl}}\bigr)^2 + 4\Delta_{a\gamma}^2}.
\]
It follows from  Eq.~\eqref{12} that $P_C(z)$ is maximized for $\beta = \pi/4$, and vanishes identically for $\beta = 0$ or $\beta = \pi/2$.

For an initially circularly polarized photon state, the analytic expression for 
$P_C(z)$ is
\begin{equation}
P_C(z) = \frac{\cos^2\!\theta\;A_{12}' + \sin^2\!\theta\;A_{13}'}
{A_0 + A_{23}'\cos\phi_{23}},
\label{13}
\end{equation}
with
\[
A_{12}' = \frac{1}{2}(C_R^2 - C_L^2)\cos\phi_{12}, 
\]
\[
A_{13}' = \frac{1}{2}(C_R^2 - C_L^2)\cos\phi_{13}, 
\]
\[
A_{23}' = \frac{1}{2}\sin^2\!\theta\cos^2\!\theta\,(C_R - C_L)^2,
\]
\[
A_0 = \frac{1}{4}\Big[(C_R + C_L)^2 + \big(1 - \tfrac{1}{2}\sin^2 2\theta\big)(C_R - C_L)^2\Big].
\]
Here $\theta$ is the photon-axion mixing angle, which characterizes the 
coupling strength and satisfies
\begin{equation}
\tan 2\theta = \frac{2\Delta_{a\gamma}}{\Delta_a - \Delta_{\mathrm{pl}}}.
\end{equation}
Considering a purely right-circularly polarized initial state ($C_R = 1$, $C_L = 0$), Eq.~\eqref{13}  can be rewritten as
\begin{equation}
P_C(z) = \frac{\cos^2\!\theta\,\cos\phi_{12} + \sin^2\!\theta\,\cos\phi_{13}}
{1 - \sin^2\!\theta\cos^2\!\theta 
+ \sin^2\!\theta\cos^2\!\theta\,\cos\phi_{23}}.
\label{eq:PC_circular}
\end{equation}
Defining the detuning parameter as  $\Delta = \Delta_a - \Delta_{\mathrm{pl}}$, Eq.~\eqref{eq:PC_circular} can be further analyzed  in three 
representative limiting regimes, according to the relative magnitude of the coupling
strength and the detuning.

\begin{figure*}[!htb]
    \centering
    \includegraphics[width=0.85\textwidth]{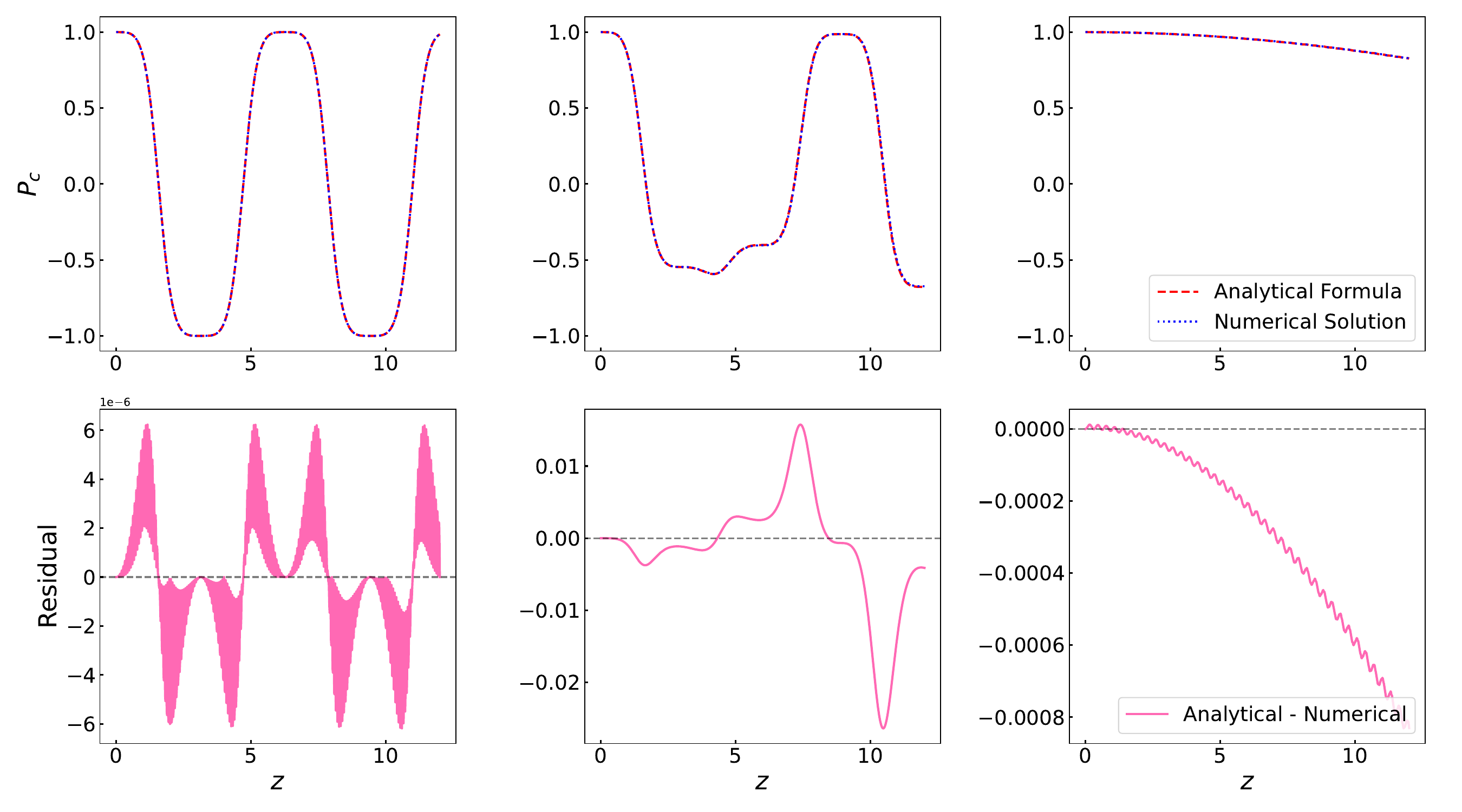}
    \caption{Comparison of the analytic and numerical solutions for the circular
polarization degree $P_C(z)$ in the three limiting cases (resonance, strong-coupling, and weak-coupling, from left to right). The upper panels show $P_C(z)$ as a function of propagation distance, while the lower panels display the corresponding residuals. Red dashed curves denote the analytic results and blue dotted curves denote the fourth-order Runge-Kutta numerical solutions. All quantities are expressed in normalized units with $\Delta_{a\gamma}=1$, and the propagation distance $z$ is in units of $\Delta_{a\gamma}^{-1}$. The initial state is purely right-handed circular polarization ($C_R=1$, $C_L=0$). The detuning $\Delta\equiv\Delta_a-\Delta_{\mathrm{pl}}$ is set to $\Delta=0$ (resonance, left panel, $\Delta_{a\gamma}/|\Delta|\to\infty$), $\Delta=0.667$ (strong coupling, middle panel, $\Delta_{a\gamma}/|\Delta|=1.5$), and $\Delta=20$ (weak coupling, right panel, $\Delta_{a\gamma}/|\Delta|=0.05$).}
    \label{fig:comparison}
\end{figure*}

{\bf Resonance:} When the axion mass term exactly cancels the
plasma effect, i.e. $\Delta_a = \Delta_{\mathrm{pl}}$, the system is at resonance. 
The mixing angle reaches its maximum value $\theta = \pi/4$, and the oscillation 
frequency reduces to $\Delta_{\mathrm{osc}} = 2\Delta_{a\gamma}$. The circular 
polarization degree simplifies to
\begin{equation}
P_C(z) = \frac{2\cos(\Delta_{a\gamma}z)}{1 + \cos^2(\Delta_{a\gamma}z)}.
\end{equation}
The corresponding evolution is shown in the left panel of Fig.~\ref{fig:comparison}. At resonance, energy is exchanged between photons and axions with maximum 
efficiency. The circular polarization degree undergoes complete periodic 
oscillations with an amplitude reaching unity, implying  that the photon polarization 
state undergoes a full reversal between pure right-circularly and pure left-circularly 
polarization. This behavior is consistent with the standard two-level Rabi oscillation picture~\cite{Yao:2026yez}.

{\bf Strong coupling:} When the coupling term dominates but the
detuning is not entirely negligible, i.e.
$2\Delta_{a\gamma} \gg |\Delta|$, the system is in the 
strong-coupling regime. The mixing angle remains close to $\pi/4$, and the 
oscillation frequency is approximated by
\begin{equation}
\Delta_{\mathrm{osc}} \approx 2\Delta_{a\gamma} 
+ \frac{\Delta^2}{4\Delta_{a\gamma}}.
\end{equation}
Substituting this approximate expression into the analytic solution, 
the phase differences become
\begin{equation}
\begin{gathered}
\phi_{12} \approx \left(\frac{1}{2}\Delta 
- \Delta_{a\gamma} 
- \frac{\Delta^2}{8\Delta_{a\gamma}}\right)z, \\
\phi_{13} \approx \left(\frac{1}{2}\Delta
+ \Delta_{a\gamma} 
+ \frac{\Delta^2}{8\Delta_{a\gamma}}\right)z, \\
\phi_{23} \approx \left(2\Delta_{a\gamma} 
+ \frac{\Delta^2}{4\Delta_{a\gamma}}\right)z.
\end{gathered}
\end{equation}
The corresponding evolution is shown in the middle panel of Fig.~\ref{fig:comparison}. In this regime, the polarization oscillations retain a large amplitude, 
while their frequency and phase are modulated by the detuning, leading to a slight 
distortion of the oscillation envelope.

{\bf Weak coupling:} When the detuning greatly exceeds the coupling
term, i.e. $2\Delta_{a\gamma} \ll |\Delta|$, the system 
enters  the weak-mixing regime. The mixing angle is small, 
$\theta \approx \Delta_{a\gamma}/\Delta \ll 1$, and 
the oscillation frequency is approximated by 
\begin{equation}
\Delta_{\mathrm{osc}} \approx \bigl|\Delta\bigr| 
+ \frac{2\Delta_{a\gamma}^2}{\bigl|\Delta\bigr|}. 
\end{equation}
Under these conditions, the circular polarization degree reads
\begin{equation}
P_C(z) \approx 
\frac{
(1 - \theta^2)\cos\!\left(\dfrac{\Delta_{a\gamma}^2}{\Delta}\,z\right)
+ \theta^2\cos\!\left[\left(\Delta 
+ \dfrac{\Delta_{a\gamma}^2}{\Delta
}\right)z\right]
}{
1 - \theta^2 + \theta^2\cos\!\left[\left(\Delta
+ \dfrac{2\Delta_{a\gamma}^2}{\Delta}\right)z\right]
}.
\end{equation}
The corresponding evolution is shown in the right panel of Fig.~\ref{fig:comparison}.
In this regime, the photon-axion conversion efficiency is strongly suppressed, and the circular 
polarization signal manifests as small-amplitude, high-frequency oscillations superimposed on a slowly varying envelope. This result is consistent with that 
obtained from second-order perturbation theory in the weak-mixing limit (see appendix \ref{B} for more details).

To verify the reliability of the analytic results, we also solve the evolution 
equations numerically and compare the solutions with the analytic expressions in 
the three limiting cases discussed above. Fig.~\ref{fig:comparison} shows the 
evolution of the circular polarization degree $P_C(z)$ as a function of propagation 
distance $z$ under the resonance, strong-coupling, and weak-coupling conditions. 
The analytic results (red dashed lines) are in excellent agreement with the 
fourth-order Runge-Kutta numerical solutions (blue dotted lines) in all cases, with maximum residuals of $6.23\times10^{-6}$ at resonance, $2.64\times10^{-2}$ 
in the strong-coupling regime, and $8\times10^{-4}$ in the weak-coupling regime.

Prior to this work, Yao et al. derived an analytic 
expression for the circular polarization degree in the weak-coupling approximation 
based on a perturbative expansion~\cite{Yao:2022col}. Masaki et al. obtained 
an analytic solution for an initially linearly polarized beam by diagonalizing the 
mixing matrix~\cite{Masaki:2017aea}. However, both of these works employed the conventional linear 
polarization basis and therefore did not directly expose the relative phase evolution of the left- and right-circular polarization modes induced by axion-photon mixing. In the present work, we 
introduce the chiral basis and carry out a direct diagonalization of the mixing matrix. This approach makes the chiral phase structure explicit and provides a compact description of the circular-polarization evolution in the resonant, strong-mixing, and weak-mixing limits within the single-domain setup considered here.

\section{Energy Dependence Characteristics of Circular Polarization Degree in Astrophysical Environments}
\label{sec:three}
Based on the analytic model developed in the preceding section, we investigate the characteristics of axion-induced circular polarization signals as photons propagate through various astrophysical environments. We first evaluate the energy dependence of photon-axion mixing efficiency across four representative magnetic  environments--the active galactic nuclei (AGN) jet, the intracluster magnetic field (ICMF), the intergalactic magnetic field (IGMF), and the Galactic magnetic field (GMF)--to identify the energy bands in which 
axion-induced circular polarization may produce potentially observable features. Subsequently, we compute the 
energy dependence of the photon circular polarization degree in the single-domain 
magnetic field model for each of these environments.

The efficiency of photon-axion mixing is governed by the physical conditions of the 
propagation medium, characterized primarily by the magnetic field strength $B$, the electron 
number density $n_e$, and the magnetic field coherence length $L$. To compare 
the mixing effects across distinct astrophysical contexts, we select four representative 
environments whose parameters are summarized in Table~\ref{tab:environments}.

\begin{table}[htbp]                                                       
\centering
\begin{tabular*}{\columnwidth}{@{\extracolsep{\fill}}cccc}
\hline
Environment & $B$ ($\mu\mathrm{G}$) & $n_e$ (cm$^{-3}$) & $L$ (kpc) \\
\hline
AGN  & $10^{6}$  & $10^{2}$  & $10^{-3}$ \\
ICMF & $1$       & $10^{-3}$ & $10$      \\
IGMF  & $10^{-3}$ & $10^{-7}$ & $5\times 10^{4}$ \\
GMF  & $1$       & $10^{-1}$ & $10^{-2}$ \\
\hline
\end{tabular*}
\caption{Fiducial parameters adopted for the four representative astrophysical 
environments considered in this work. The representative parameter values are compiled from~\cite{Yao:2022col,Ghisellini:2009fj}.}
\label{tab:environments}
\end{table}
The strength of photon-axion mixing can be characterized by the mixing angle
\begin{figure*}[!htb]
    \centering
    \includegraphics[width=0.9\textwidth]{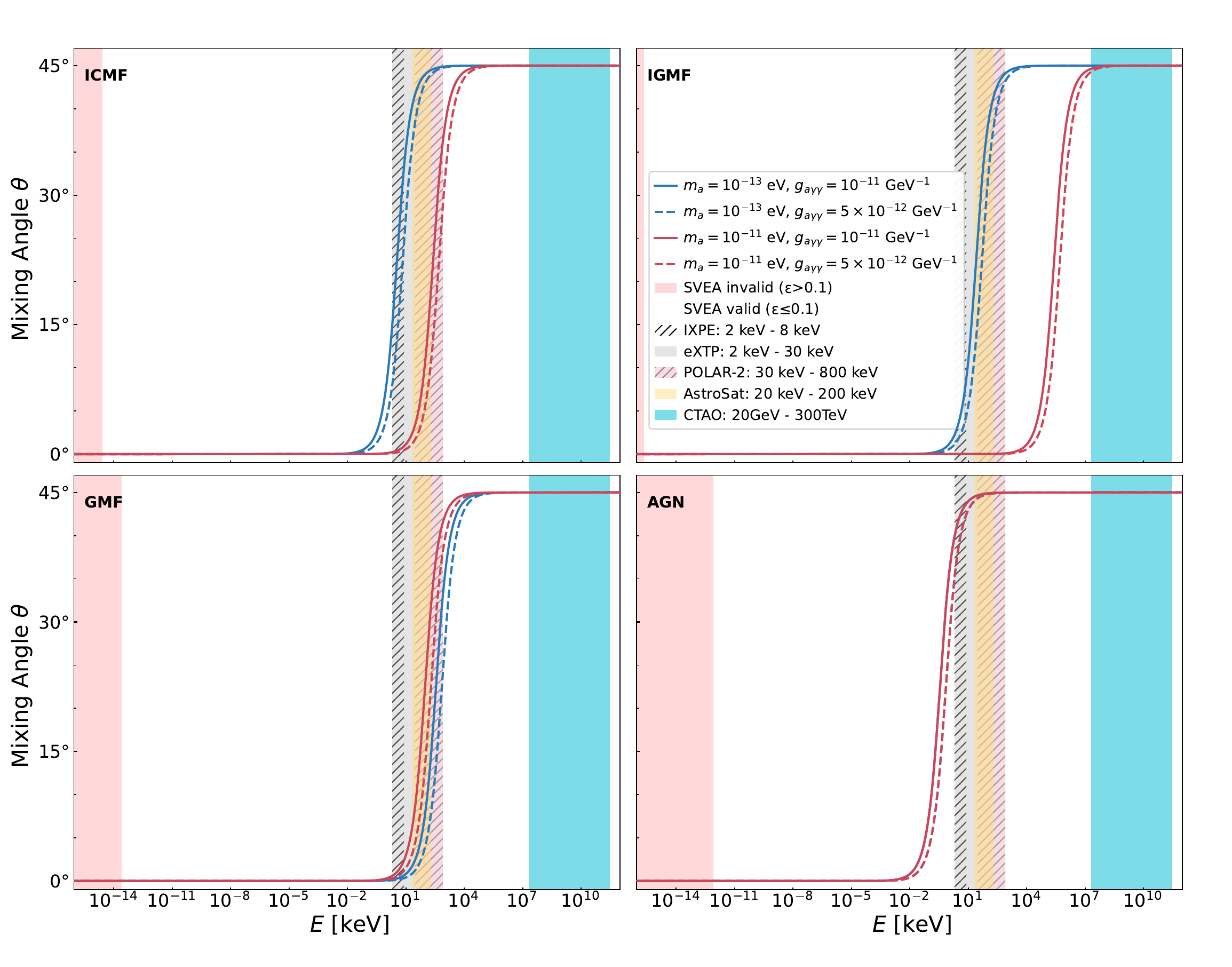}
    \caption{Mixing angle as a function of photon energy in four representative 
astrophysical environments. Each panel corresponds to one environment (ICMF, IGMF, GMF, and AGN), and each 
curve corresponds to a different set of axion parameters. The red shaded region on 
the left marks the energy range where the SVEA may fail. The colored shaded 
regions on the right indicate the approximate energy coverage of different telescopes (IXPE~\cite{ratheesh2021imaging, Soffitta:2013hla}, 
eXTP~\cite{Wang:2025hyu}, POLAR-2~\cite{Kole:2024ekt}, AstroSat~\cite{Bhalerao:2016lex}, and CTAO~\cite{Mazin:2019ykz, Lopez-Oramas:2025vld}).}
    \label{fig:mixing_angle}
\end{figure*}
\begin{equation}
\theta = \frac{1}{2}\arctan\left( \frac{2\Delta_{a\gamma}}{\Delta_{a} - \Delta_{pl}} \right) = \frac{1}{2}\arctan\left( \frac{2g_{a\gamma\gamma}B_{T}E}{m_{a}^{2} - \omega_{pl}^{2}} \right)
\label{eq:mixing_angle}
\end{equation}
Fig.~\ref{fig:mixing_angle} shows the mixing angle as a function of photon energy 
in each of the four astrophysical environments considered above, computed for four 
representative sets of axion parameters. At radio and optical energies, 
all curves approach $\theta = 0$, indicating that the system resides in the 
weak-coupling regime where photon-axion mixing is highly inefficient. As the photon 
energy enters the X-ray band, the mixing angle begins to deviate appreciably from 
zero and increases with energy. At high energies (GeV--TeV band), $\theta$ 
asymptotically approaches $\pi/4$, signaling the onset of the resonant 
strong-coupling regime. These results imply that the axion-induced circular polarization 
degree varies with photon energy, and most pronounced energy dependence, for the representative parameters shown here, occurs from the X-ray to MeV band. To indicate the validity of the theoretical framework,  we also show the range of applicability of the slowly varying envelope approximation (SVEA) in the figure. This approximation requires 
that the scale over which the matrix elements of the evolution Hamiltonian vary be much larger 
than the photon wavelength, a condition that can be quantified by the dimensionless parameter
\begin{equation}
    \epsilon = \max\!\left(\frac{g_{a\gamma\gamma}B_T}{2k},\,
               \frac{\left|m_a^2 - \omega_{\rm pl}^2\right|}{2\omega k}\right) \ll 1,
    \label{eq:SVEA}
\end{equation}
where $k \simeq \omega$ is the photon wavenumber. The low-energy region in Fig.~\ref{fig:mixing_angle} for which $\epsilon \gtrsim 0.1$, namely the region where the SVEA 
may break down, is excluded the subsequent analysis.

\begin{figure*}[htbp]
    \centering
    \includegraphics[width=0.75\textwidth]{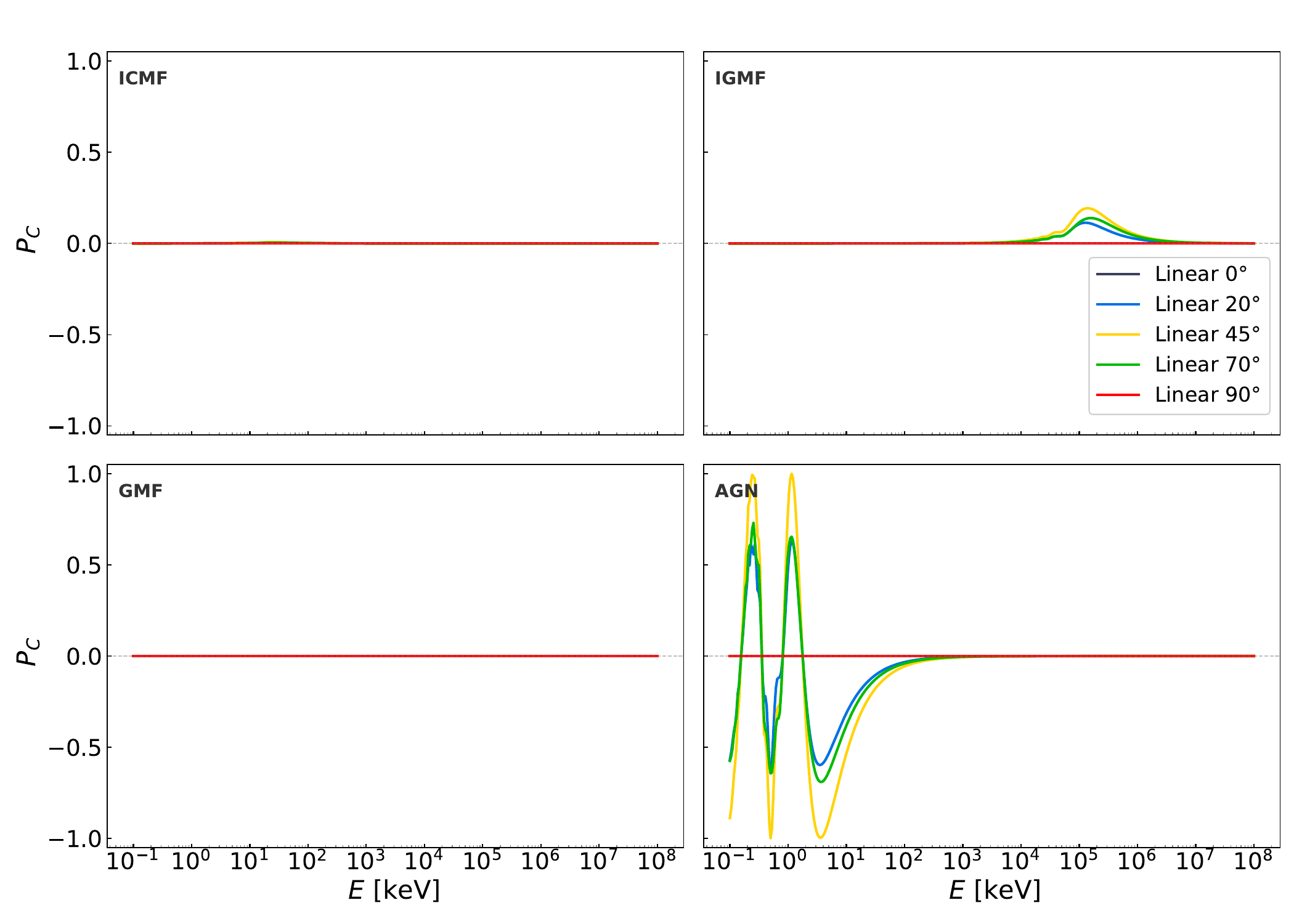}
    \caption{Circular polarization degree as a function of photon energy for different 
initial linear polarization angles in the single-domain magnetic-field model. Each panel corresponds 
to one astrophysical environment (ICMF, IGMF, GMF, and AGN), and each curve corresponds to a different initial 
polarization angle $\beta$.}
    \label{fig:single_domain_pure}
\end{figure*}

\begin{figure*}[htbp]
    \centering
    \includegraphics[width=0.75\textwidth]{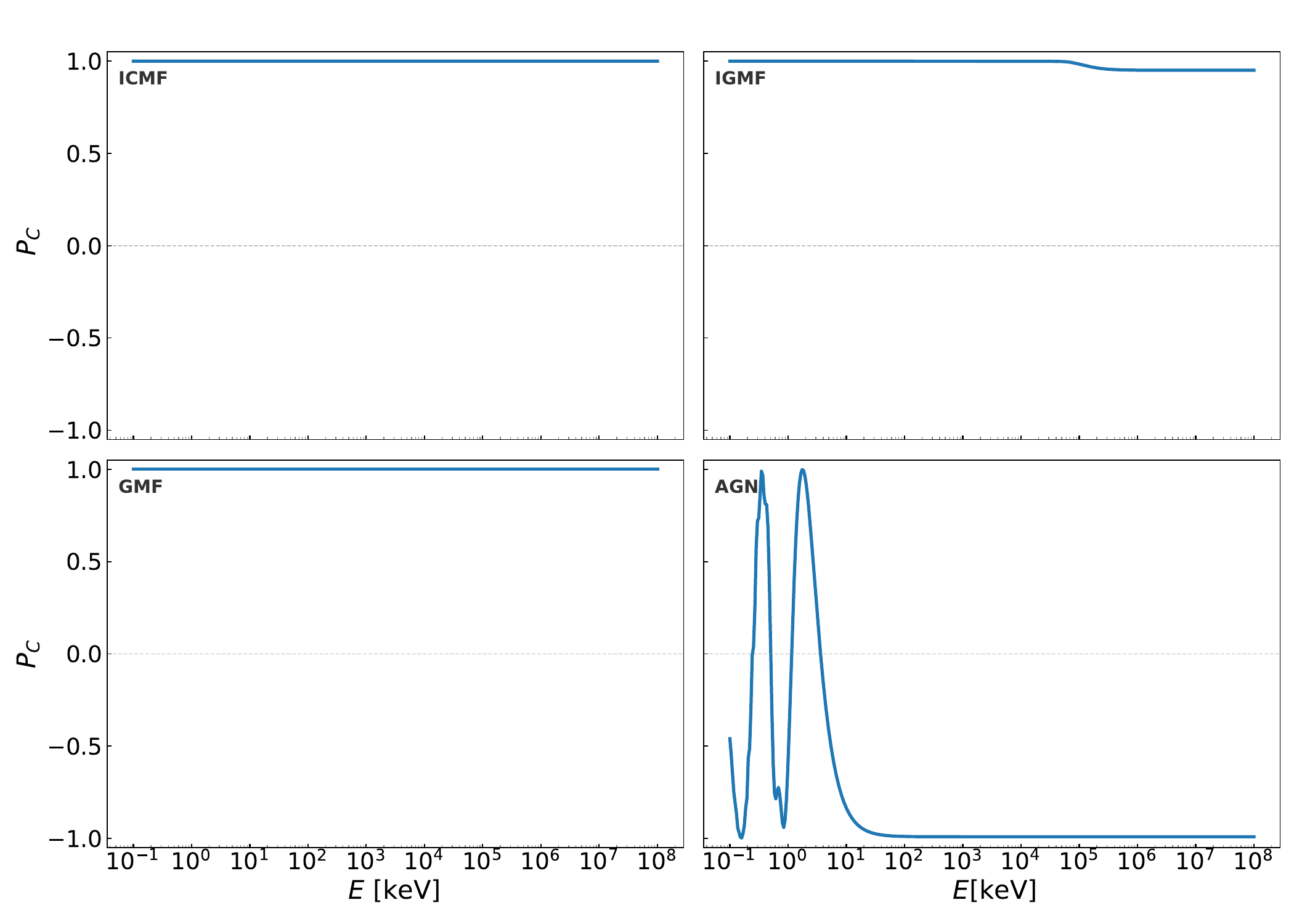}
    \caption{Circular polarization degree as a function of photon energy for an initially 
purely right-handed circularly polarized state in the single-domain model. Each panel 
corresponds to one astrophysical environmentent (ICMF, IGMF, GMF, and AGN).}
    \label{fig:single_domain_mixed}
\end{figure*}

The photon-axion coupling strength exhibits a strong energy dependence. 
Motivated by this property, we compute the circular polarization degree as a 
function of photon energy for propagation through a single homogeneous magnetic domain in 
each of the four environments. In the single-domain model, the magnetic field 
strength and orientation are held constant, and the propagation distance is taken to 
be the magnetic coherence length $L$. We adopt a representative set of axion 
parameters, $m_a = 1\times10^{-11}\ \mathrm{eV}$ and 
$g_{a\gamma\gamma} = 1\times10^{-11}\ \mathrm{GeV}^{-1}$, and consider two 
initial polarization states, namely a purely linearly polarized state 
(Fig.~\ref{fig:single_domain_pure}) and a purely right-handed circularly polarized state (Fig.~\ref{fig:single_domain_mixed}).

Fig.~\ref{fig:single_domain_pure} shows the results for an initially linearly 
polarized beam. The circular polarization degree $P_C$ depends strongly on the 
initial linear polarization angle $\beta$. In the ICMF and GMF, $P_C$ remains negligibly small for all values 
of $\beta$. This is because both environments possess relatively weak magnetic fields 
($B\sim 1\,\mu\mathrm{G}$) and coherence lengths $L$ much shorter than the 
oscillation length $L_{\mathrm{osc}}\equiv 2\pi/\Delta_{\mathrm{osc}}$, so that the accumulated circular-polarization signal remains small. In the IGMF, $P_C$ is maximized at $\beta = \pi/4$ and vanishes at 
$\beta = 0$ and $\pi/2$, in agreement with the analytic expressions derived in 
Sec.~II. Furthermore, the long path length of the IGMF ($L\sim 50\,\mathrm{Mpc}$) 
provides sufficient propagation distance for the polarization signal to accumulate, 
resulting in a broad, slowly varying peak in $P_C$ at the MeV energy band. In the AGN jet, 
the strong magnetic field ($B\sim 10^{6}\,\mu\mathrm{G}$) enhances 
photon-axion coupling. In the X-ray band, the system lies in the off-resonance regime 
($\theta\ll\pi/4$), where the interference phase $\phi_{23}(E)$ is highly sensitive 
to energy, causing $P_C$ to oscillate rapidly with large amplitude. By contrast, 
although the system enters the strong-coupling regime at TeV energies, the phase evolution becomes less sensitive to energy, and $P_C$ consequently tends to stabilize.

Fig.~\ref{fig:single_domain_mixed} presents the results for an initially 
right-handed circularly polarized beam. In the ICMF and GMF, the variation of $P_C$ 
remains likewise negligible. In the IGMF, $P_C$ exhibits a broad, shallow dip structure. In the 
AGN jet, $P_C$ undergoes large-amplitude oscillations with an amplitude spanning the full 
range $[-1,\,1]$, indicating that the polarization state can undergo nearly complete 
conversion between left- and right-handed circular polarization as a function of photon energy. This distinctive energy-dependent signature could provide a useful diagnostic for identifying axion-induced circular-polarization features.

\section{Evolution of Circular Polarization in Multi-Domain Propagation and Constraints on Axion Parameters}
\label{sec:four}
In realistic astrophysical environments, magnetic field structures typically 
exhibit complex inhomogeneous properties, which differ significantly from the 
single-domain uniform magnetic field adopted in Section~III. To describe the propagation of photons through real astrophysical medium, we study the evolution of axion-induced circular polarization 
under multi-domain propagation. We construct a three-zone line-of-sight propagation model that includes the AGN jet, IGMF, and GMF. Taking the blazar S4~0954+65 as a representative 
source, we numerically simulate the propagation from the source to the observer, and analyze the evolution of the circular polarization degree as a 
function of redshift and photon energy. We then combine the result with the existing upper limit on the optical circular polarization of this source to constrain the axion parameter space.
\subsection{Multi-Domain Magnetic Field Configuration and Numerical Simulation Method}
We consider high-energy photons from the blazar S4~0954+65 propagating along the line of sight through three magnetized regions in sequence, namely the AGN, the IGMF, and the GMF.
The source parameters are right ascension $09^h58^m47.24^s$, declination 
$+65^\circ33'54.81''$, and redshift $z = 0.367$~\cite{Liodakis:2023qnt}.

For the AGN jet, we adopt an axisymmetric helical magnetic-field model. Following the Clausen--Brown model, we decompose the 
magnetic field in cylindrical coordinates $(r,\phi,z)$ into an axial component 
$B_z$ and a toroidal component $B_\phi$, whose radial profiles are given by 
Bessel functions~\cite{Clausen-Brown:2011upb}:
\begin{equation}
\begin{aligned}
B_z(r,z) &= B_T^{\rm jet}(z)\,\frac{J_0(k\rho)}{\sqrt{J_0^2 + J_1^2}}, \\
B_\phi(r,z) &= B_T^{\rm jet}(z)\,\frac{J_1(k\rho)}{\sqrt{J_0^2 + J_1^2}},
\end{aligned}
\end{equation}
where $\rho = r/R_{\rm jet}$, and $k = 2.405$ is the first zero of $J_0$, 
ensuring $B_z = 0$ at the jet boundary. Here $B_T^{\rm jet}(z)$ describes the 
overall evolution of the magnetic field strength with distance and serves as a 
unified normalization factor for all components. The magnetic field phase is
\begin{equation}
\phi_{\rm rot}(z) = \phi_0 + 10\,\ln\left(\frac{z}{z_E}\right),
\end{equation}
where $\phi_0 \in [0,2\pi]$ and $z_E = 0.01\,{\rm pc}$. Taking into account the 
viewing angle of approximately $5.5^\circ$ between the jet axis and the line of 
sight~\cite{Jorstad:2017bga,volvach2016non}, we rotate the magnetic field from the jet frame to the observer frame and 
then project it to obtain the transverse component. The overall magnetic field strength and electron number density follow power-law profiles~\cite{Meyer:2014epa, Tavecchio:2014yoa}, as
\begin{equation}
\begin{aligned}
B_T^{\rm jet}(z) &= B_{T,0}\left(\frac{z}{z_E}\right)^{-1}, \\
n_e^{\rm jet}(z) &= n_{e,0}\left(\frac{z}{z_E}\right)^{-2},
\end{aligned}
\end{equation}
where $B_{T,0} \sim 1\,{\rm G}$ and $n_{e,0} \sim 10^2\,{\rm cm}^{-3}$. The jet 
region is discretized into approximately $150$--$200$ domains. In each domain, the magnetic-field strength and geometry are determined by the model above, while the field orientation varies continuously across domain boundaries. 

For the IGMF, we adopt a random magnetic domain model to characterize the 
turbulent magnetic field~\cite{Meyer:2021pbp}. The propagation path is divided into a series of 
cells, each with a coherence length of $L_{\rm coh} \sim 1\ \mathrm{Mpc}$. 
Within each domain, the magnetic field strength $B_{\rm IGMF} \sim 1\ \mathrm{nG}$ 
is held constant, while the field orientation is randomly and uniformly 
distributed over the interval $[0, 2\pi]$. The electron number density is 
assigned a typical value of $n_e^{\rm IGM} \sim 10^{-7}\ \mathrm{cm}^{-3}$. 
The total number of IGMF magnetic domains is determined by the source redshift. 
For a source at $z = 0.367$, the comoving distance is approximately 
$1500\ \mathrm{Mpc}$, corresponding to roughly $1500$ domains.

For the GMF, we adopt the Jansson--Farrar  model~\cite{Jansson:2012pc}, which is based on a joint fit to approximately $4\times10^4$ Faraday rotation 
measurements of extragalactic radio sources and pulsars, together with the 
WMAP synchrotron polarization data. This model provides a parametric 
description of the Galactic disk field, the halo field, and the out-of-plane 
field components. The distribution of the electron number density 
$n_e$ is described by the Cordes--Lazio model~\cite{cordes2002ne2001}.

For the three regions described above,  we generate the corresponding 
sequence of magnetic domains. The evolution of the photon-axion system within 
each domain is governed by  Eq.~\eqref{eq:evolution}. The state vector 
$\psi = (|R\rangle, |L\rangle, |a\rangle)^{\rm T}$ evolves within each domain 
as $\psi(z) = \mathcal{U}(z)\psi(0)$, where 
$\mathcal{U}(z) = \exp(-i\mathcal{H}_{\rm chiral}z)$, and 
$\mathcal{H}_{\rm chiral}$ is the mixing matrix given in Sec.~\ref{sec:Analytic model}. For a propagation path consisting of $N$ magnetic domains, the 
overall evolution operator is the ordered product of the individual domain 
operators,
\begin{equation}
\psi_{\rm final} = \mathcal{U}_N(z_N) \cdots \mathcal{U}_2(z_2) 
\mathcal{U}_1(z_1)\, \psi_{\rm initial},
\end{equation}
where $\mathcal{U}_i(z_i)$ denotes the evolution operator of the $i$-th domain. 
In the numerical calculation, the state vector is updated iteratively domain by 
domain as $\psi_{i+1} = \mathcal{U}_i(z_i)\,\psi_i$ until the final state is 
obtained. Given that the initial polarization state of an astrophysical source 
may be a mixture of linear polarization and unpolarized components, we employ a Monte Carlo 
ensemble method to compute the final ensemble-averaged circular polarization 
degree. Optical observations of the blazar S4~0954+65 show a linear polarization
degree of $\Pi_l = 30.9\%$ with an electric-vector position angle (EVPA) of $83^\circ$~\cite{Liodakis:2023qnt}, and we construct a photon ensemble of $2000$
initial pure states to simulate its polarization properties. Among these, $30.9\%$
of the samples are assigned the same linearly polarized state (with polarization angle fixed at the observed EVPA), while the
remaining $69.1\%$ are generated by randomly sampling pure states with uniformly
random phases and polarization angles, which represents the
unpolarized component. The multi-domain iterative evolution is performed 
independently for each initial state in the ensemble, and the circular 
polarization degree $\langle P_C \rangle$ is obtained by statistically averaging 
over all final states.

\subsection{Redshift Evolution and Energy-Dependent Features}

\begin{figure}[htbp]
\centering
\includegraphics[width=1\columnwidth]{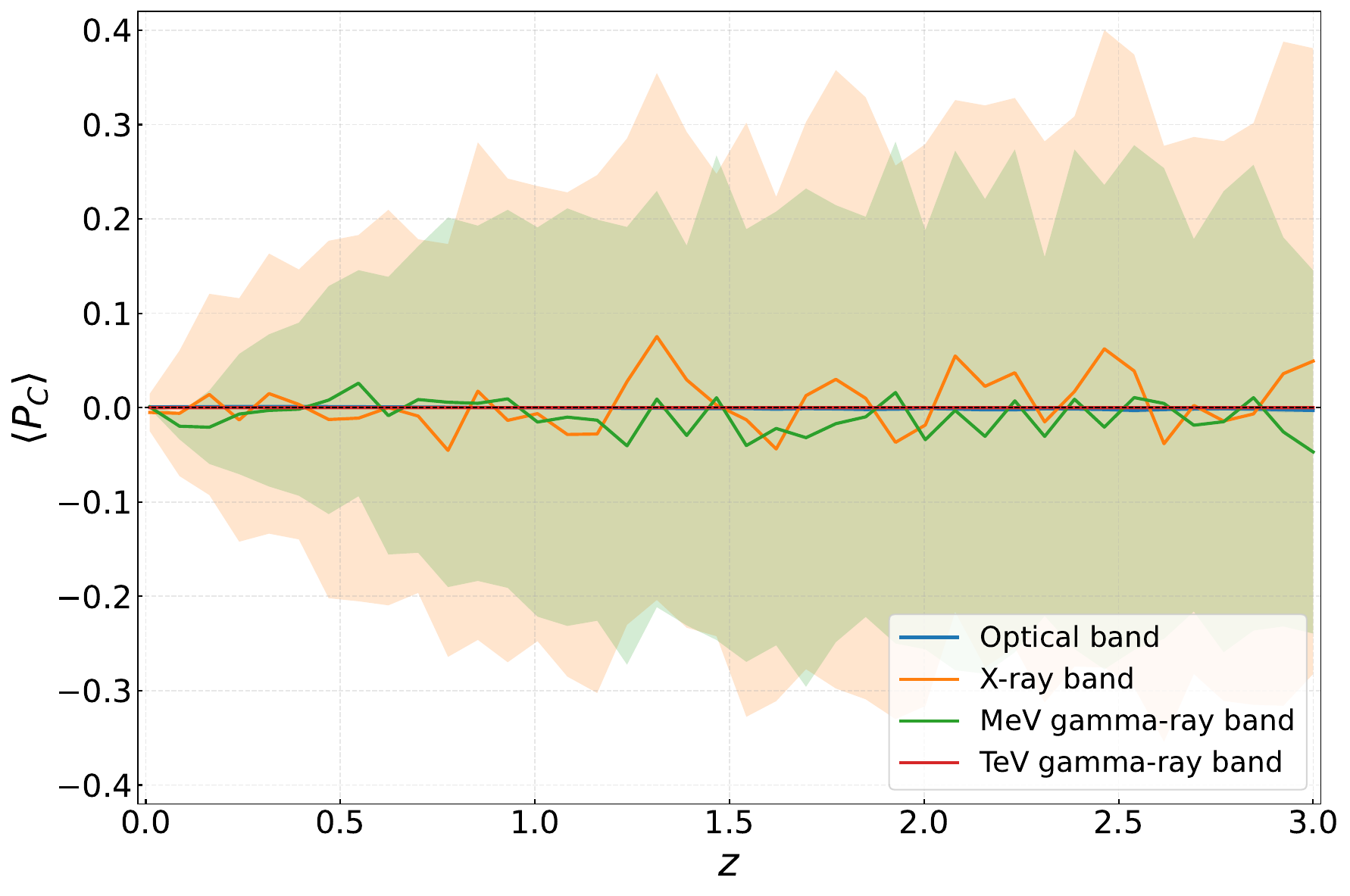}
\caption{Average circular polarization degree $\langle P_C \rangle$ as a function of source redshift $z$ under multi-domain propagation for an ALP with $m_a = 10^{-13}\ \mathrm{eV}$ and $g_{a\gamma\gamma} = 1\times10^{-11}\ \mathrm{GeV}^{-1}$. Blue, orange, green, and red lines denote the optical, X-ray, MeV gamma-ray, and TeV gamma-ray bands, respectively.}
\label{fig5}
\end{figure}

Using the multi-domain magnetic field model described above, we compute the 
ensemble-averaged circular polarization degree as a function of source redshift $z$. Fig.~\ref{fig5} shows the 
ensemble-averaged circular polarization degree $\langle P_C \rangle$ as a 
function of source redshift $z$ for the optical, X-ray, MeV gamma-ray, and TeV 
gamma-ray bands, computed for axion parameters $m_a = 10^{-13}\ \mathrm{eV}$ 
and $g_{a\gamma\gamma} = 1\times10^{-11}\ \mathrm{GeV}^{-1}$. The calculation 
assumes that photons traverse the entire propagation medium at their 
emission-frame energy.  In the figure, the solid curves represent the ensemble mean obtained 
from multi-domain evolution, and the shaded regions indicate the $1\sigma$ 
uncertainty. The results show that in the X-ray and MeV bands, the circular 
polarization degree exhibits stochastic fluctuations with increasing 
redshift, arising from phase accumulation as photons propagate through 
multiple magnetic domains. As the redshift $z$ increases, photons traverse a 
larger number of IGMF domains. Since azimuthal angle of the transverse magnetic field within each domain is randomly distributed, random phase 
differences are continuously introduced during the evolution. This random 
superposition of interference phases across different domains causes the net 
polarization degree $\langle P_C \rangle$ to fluctuate irregularly about zero 
with redshift, while the $1\sigma$ uncertainty broadens with propagation 
distance following the statistics of a random walk. This redshift-dependent feature carries clear observational implications. From 
a theoretical standpoint, the intrinsic polarization properties of astrophysical sources are unlikely to naturally produce such pronounced fluctuations spanning a wide range of redshifts. Thus, comparing circular polarization measurements for AGN at different redshifts could help separate the propagation-induced photon-axion mixing signal from the source-intrinsic contribution, thereby providing an independent probe of the axion parameter space.

\begin{figure}[htbp]
\centering
\includegraphics[width=1\columnwidth]{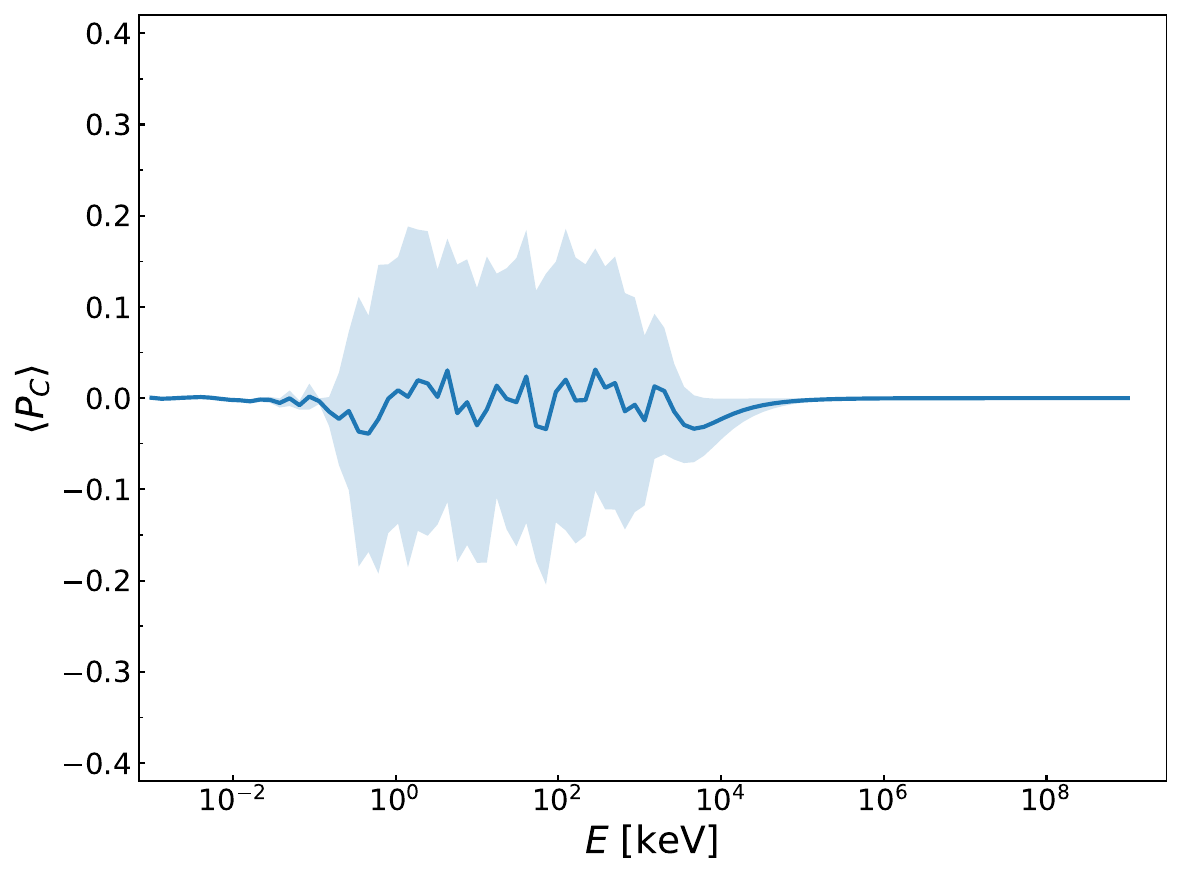}
\caption{Average circular polarization degree $\langle P_C \rangle$ as a function of photon energy $E$ under multi-domain propagation for an ALP with $m_a = 10^{-13}\ \mathrm{eV}$ and $g_{a\gamma\gamma} = 1\times10^{-11}\ \mathrm{GeV}^{-1}$.}
\label{fig6}
\end{figure}

Furthermore, we investigate the energy dependence of the circular polarization degree within the multi-domain model at a fixed source redshift of $z = 0.367$, as shown in Fig.~\ref{fig6}. The results show a clear energy dependence. In the X-ray to MeV band 
($E \sim 1$--$10^{4}\ \mathrm{keV}$), the ensemble-averaged circular 
polarization degree $\langle P_C \rangle$ exhibits large stochastic 
fluctuations, whereas the amplitude of the fluctuations is suppressed at lower energies (radio to optical) and at very high energies. This 
non-trivial energy-dependent feature provides a key observational test for axion searches. Observationally, upper limits on the circular polarization 
degree from a single waveband (e.g., optical or X-ray) already allow independent constraints on the axion parameter space. Looking 
ahead, broadband multi-wavelength circular polarization observations spanning a 
wide energy range, particularly in the X-ray to MeV band, will more 
effectively break the physical degeneracy between the photon--axion mixing 
propagation effect and intrinsic emission mechanisms.

\subsection{Constraints on Axion Parameters from S4~0954+65}
\begin{figure}[htbp]
\centering
\includegraphics[width=1\columnwidth]{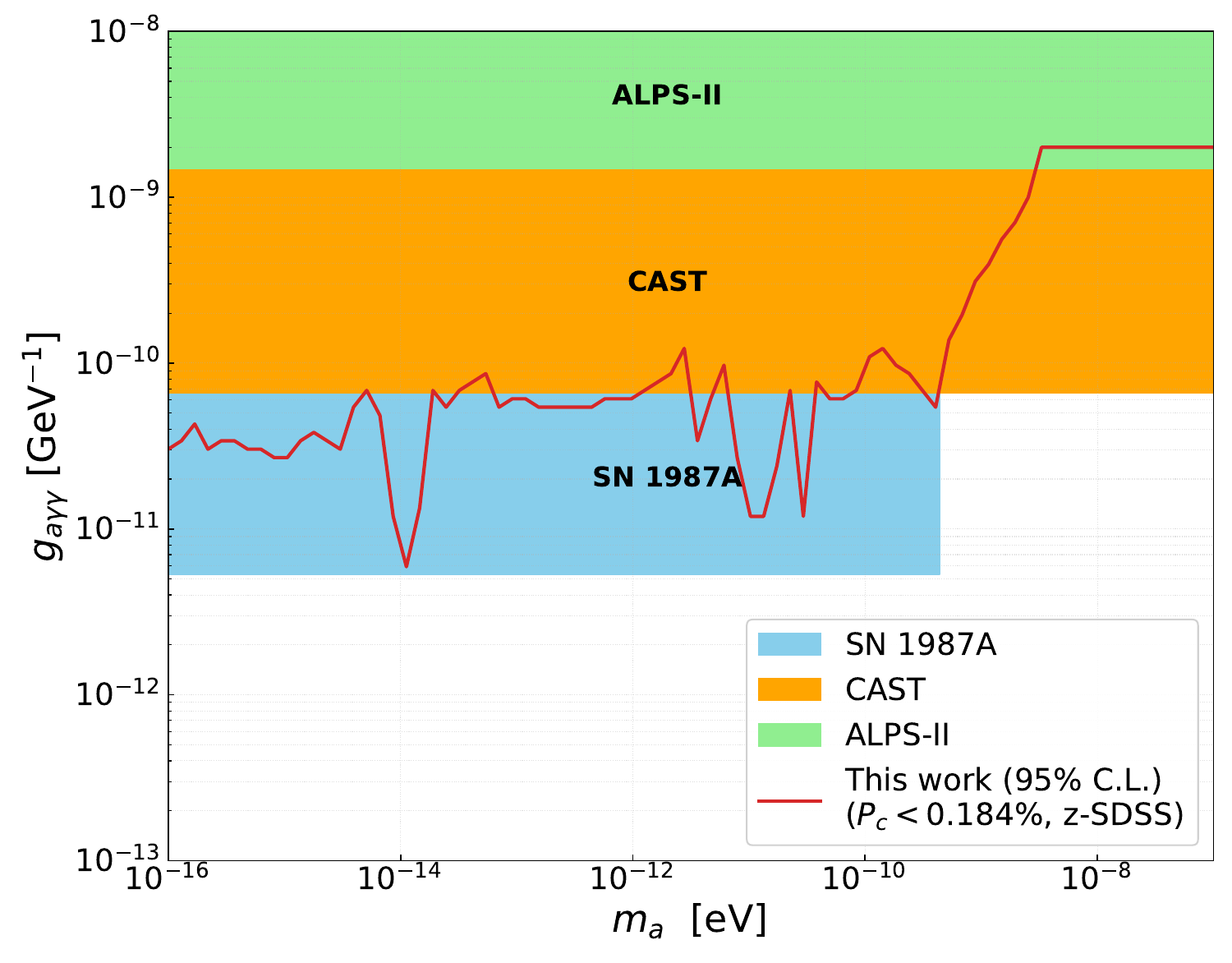}
\caption{Constraints on the axion parameter space derived from the upper limit
on circular polarization of the blazar S4~0954+65. The red solid curve
represents the exclusion boundary obtained in this work based on $P_c < 0.184\%$
in the z-SDSS band ($E\simeq1.36\,\mathrm{eV}$), at the $95\%$ confidence level (i.e. $|P_C|>0.184\%$ in at least $95\%$ of the random magnetic-field realizations) (the region above the curve is excluded). For comparison, the blue, orange, and
green shaded regions represent previously obtained bounds from SN~1987A, CAST,
and ALPS-II, respectively.}

\label{fig:parameter_exclusion}
\end{figure}

The blazar S4~0954+65 is one of the few low-synchrotron-peaked blazars detected
at TeV gamma-ray energies, exhibiting a high degree of optical linear
polarization, with an observational upper limit on the optical circular
polarization of $P_c < 0.184\%\ (3\sigma)$~\cite{Liodakis:2023qnt}. This upper limit was obtained in the z-SDSS broad-band filter ($\Delta\lambda\simeq830$--$1000$~nm), corresponding to a central frequency $\nu_{\rm obs}\simeq3.28\times10^{14}\,\mathrm{Hz}$, i.e. a photon energy $E\simeq1.36\,\mathrm{eV}$. Based on this upper limit, we
derive constraints on the axion parameter space. Because the predicted circular polarization degree in the multi-domain model depends on the specific stochastic realization of the random magnetic fields (cf. Fig.~\ref{fig6}), for a fixed point in the ($m_a$, $g_{a\gamma\gamma}$) plane the computed $P_C$ fluctuates from one field realization to another and can lie either above or below the observational upper limit. We therefore formulate the exclusion statistically. For each point in the ($m_a$, $g_{a\gamma\gamma}$) plane, we generate $300$ independent random magnetic-field realizations. For each realization, we compute the ensemble-averaged circular polarization degree at the observed energy $E\simeq1.36\,\mathrm{eV}$ using the multi-domain propagation model described above. A parameter point is excluded at the $95\%$ confidence level if $|P_C|>0.184\%$ in at least $95\%$ of the magnetic-field realizations. The resulting exclusion region is shown in Fig.~\ref{fig:parameter_exclusion}. For $m_a \sim 10^{-16}$--$10^{-10}\ \mathrm{eV}$, we
constrain
$g_{a\gamma\gamma} \lesssim 3\times 10^{-11}\ \mathrm{GeV}^{-1}$ (95\% C.L.), with the most
stringent bound occurring near $m_a \sim 10^{-14}\ \mathrm{eV}$, corresponding
to $g_{a\gamma\gamma} \sim 6\times 10^{-12}\ \mathrm{GeV}^{-1}$. Over $m_a \sim 10^{-16}$--$10^{-10}\ \mathrm{eV}$, this exclusion is broadly comparable to the CAST bound and, at the most sensitive masses, reaches the level of the SN~1987A limit, while lying well below the projected ALPS-II sensitivity. Being based on circular polarization, it provides a complementary constraint.

Constraining the axion parameter space via astrophysical energy spectra is another widely adopted approach. Recently, Zhu et al.\ ~\cite{Zhu:2024kmu} derived  a constraint of $g_{a\gamma\gamma} \sim 10^{-11}\ \mathrm{GeV}^{-1}$ using the 
very-high-energy gamma-ray spectrum of the Galactic center source HESS~J1745-290. Malyshev et al.\ ~\cite{Malyshev:2025iis} improved this bound to 
$g_{a\gamma\gamma} \sim 2\times 10^{-12}\ \mathrm{GeV}^{-1}$ by stacking the 
gamma-ray spectra of active galactic nuclei behind galaxy clusters. These 
results are broadly comparable in magnitude to the constraint obtained in the 
present work. However, spectral methods are sensitive to assumptions 
about the intrinsic spectral shape of the source and require modeling of 
the magnetic field and radiation background along the line of sight, leading to systematic uncertainties. Moreover, intrinsic source variability can 
produce spurious signals that may mimic photon--axion oscillations. In 
contrast, polarization-based methods are independent of the spectral shape and 
thus less affected by these spectral uncertainties. Zhang et al.\ ~\cite{Zhang:2024jvq} placed 
stringent constraints of $g_{a\gamma\gamma} \lesssim 10^{-13}\ \mathrm{GeV}^{-1}$ by using X-ray linear polarization observations of gamma-ray bursts.
Compared with the linear polarization results of Zhang 
et al., the constraint on the axion coupling constant derived in this work is 
weaker, which can be primarily attributed to systematic differences in the 
observational waveband and sample redshift. This work employs the optical band, 
whereas Zhang et al.\ utilize the X-ray band, in which the photon--axion mixing 
efficiency is substantially higher. In addition, their gamma-ray burst sample 
spans higher redshifts ($z \sim 1$), providing longer mixing path lengths. It 
is worth noting that linear polarization is susceptible to contamination from 
intrinsic polarization of the astrophysical environment. By contrast, circular 
polarization offers a distinctive advantage of intrinsically low background.
This is because standard astrophysical emission mechanisms produce negligible circular 
polarization, and any significant detection would therefore constitute a strong 
indicator of new physics. Future observations combining X-ray circular 
polarimetry with broadband, multi-redshift joint analyses are expected to 
substantially enhance the constraining power on axion parameters.

\section{Conclusion}
\label{sec:five}
This study investigates the impact of photon-axion mixing on the evolution of photon circular polarization in astrophysical environments. Within the chiral basis framework, we solve the evolution equations of the photon-axion mixing system and derive an analytic expression for the circular polarization degree $P_C(z)$, which provides a compact description of circular polarization evolution in the resonant, strong coupling, and weak coupling regimes. Specifically, at resonance, the photon polarization state undergoes nearly complete periodic reversals. In the strong coupling regime, polarization oscillations maintain substantial amplitudes, although their frequencies and phases are modulated by the detuning, resulting in slight distortions of the oscillation envelope. In the weak coupling regime, the circular polarization signal exhibits small-amplitude, high frequency oscillations superimposed on a slowly varying envelope. This analytic treatment does not rely on a weak-mixing expansion and makes explicit how the opposite chiral phases of the left- and right-handed circular polarization modes generate relative phase shifts and amplitude differences, allowing initially linearly polarized light to acquire a nonzero circular polarization component.

Building upon the analytic model, we investigate photon-axion mixing and its energy dependence in four representative astrophysical environments. The results show that mixing efficiency increases with photon energy, with the X-ray to MeV band being the key window for searching for axion-induced circular polarization signals. Single-domain magnetic field calculations show that the circular polarization degree is sensitive to energy, initial polarization angle, and environmental parameters, with its energy-dependent structure providing identifiable observational signatures for axion searches.

To more realistically describe photon propagation through astrophysical environments, we construct a multi-domain cascade model incorporating helical magnetic fields in AGN jets, intergalactic random magnetic fields, and large-scale regular Galactic magnetic fields. Using the blazar S4 0954+65 as a case study, we perform numerical simulations. The results show that phase accumulation in random magnetic domains causes the circular polarization degree to exhibit irregular fluctuations with redshift and produces pronounced energy-dependent structures in the X-ray to MeV band. These features provide potential diagnostics for distinguishing axion-induced propagation effects from intrinsic source polarization mechanisms. Based on the optical circular polarization observational upper limit \(P_C<0.184\%\) (z-SDSS band, $E\simeq1.36\,\mathrm{eV}$) from this source, we statistically constrain the axion parameter space by excluding parameter points for which $|P_C|>0.184\%$ in at least $95\%$ of the random magnetic-field realizations, obtaining an upper limit at the level of \(g_{a\gamma\gamma}\lesssim3\times10^{-11}\,\mathrm{GeV}^{-1}\) (95\% C.L.) for $m_a \sim 10^{-16}$--$10^{-10}\ \mathrm{eV}$, with the strongest constraint reaching $g_{a\gamma\gamma}\sim6\times10^{-12}\,\mathrm{GeV}^{-1}$ near \(m_a\sim10^{-14}\,\mathrm{eV}\). Compared to spectral methods and linear polarization methods, the circular polarization approach is less susceptible to astrophysical background contamination and does not depend on assumptions about the intrinsic spectrum of the source, thereby providing unique advantages for axion detection. Future circular polarization measurements in the X-ray to MeV band could further test this scenario and improve sensitivity to ultralight axions.

\section*{Acknowledgments}
We sincerely thank the anonymous referee for constructive suggestions. We are grateful for the financial support from the National
Natural Science Foundation of China (grant Nos. 12563003 and 12103022) and
the Special Basic Cooperative Research Programs of the Yunnan Provincial Undergraduate Universities Association (grant No. 202301BA070001-104). N.D.
is sincerely grateful for the financial support of the Xingdian Talents Support
Program and Yunnan Province (grant No. XDYC-QNRC-2022-0613). Y.Y.T. is
sincerely grateful for the financial support of the Xing Dingyu Academician
Workstation of Yunnan Province (202605AF350035) and the Yunnan Provincial
Foreign Talent Introduction Program (202505AP120035).

\appendix

\section{Analytic derivation of the circular-polarization degree by diagonalization}
\label{A}
We solve the evolution of the photon-axion system by diagonalizing 
the Hamiltonian in the chiral basis Eq.~\eqref{9}. Solving the secular 
equation $\det(\mathcal{H}_{\text{chiral}} - E\mathbb{I}) = 0$, we
obtain three eigenvalues:
\begin{equation}
\begin{gathered}
    E_1 = \Delta_{\mathrm{pl}}, \quad \\
    E_2 = \frac{\Delta_{\mathrm{pl}} + \Delta_a - \Delta_{\text{osc}}}{2}, \quad \\
    E_3 = \frac{\Delta_{\mathrm{pl}} + \Delta_a + \Delta_{\text{osc}}}{2},
\end{gathered}
\end{equation}
where $\Delta_{\text{osc}} = \sqrt{(\Delta_a - \Delta_{\mathrm{pl}})^2 
+ 4\Delta_{a\gamma}^2}$ is the oscillation frequency. The corresponding 
normalized eigenstates are
\begin{equation}
\begin{gathered}
|\psi_1\rangle = \frac{1}{\sqrt{2}}\big(|R\rangle + |L\rangle\big), \\
|\psi_2\rangle = N_2\left(-\frac{i\sqrt{2}\,\Delta_{a\gamma}}{B}|R\rangle 
+ \frac{i\sqrt{2}\,\Delta_{a\gamma}}{B}|L\rangle + |a\rangle\right), \\
|\psi_3\rangle = N_3\left(\frac{i\sqrt{2}\,\Delta_{a\gamma}}{B'}|R\rangle 
- \frac{i\sqrt{2}\,\Delta_{a\gamma}}{B'}|L\rangle + |a\rangle\right),
\end{gathered}
\label{eq:psi123}
\end{equation}
where
\begin{equation*}
    B = \frac{\Delta_a - \Delta_{\mathrm{pl}} - \Delta_{\text{osc}}}{2}, \quad 
    B' = \frac{\Delta_a - \Delta_{\mathrm{pl}} + \Delta_{\text{osc}}}{2},
\end{equation*}
\begin{equation*}
    N_2 = \left(1 + \frac{4\Delta_{a\gamma}^2}{B^2}\right)^{-1/2}, \quad 
    N_3 = \left(1 + \frac{4\Delta_{a\gamma}^2}{B'^2}\right)^{-1/2}.
\end{equation*}
For a general initial state 
$|\psi(0)\rangle = C_R|R\rangle + C_L|L\rangle + C_a|a\rangle$, projecting the initial state onto each eigenstate $|\psi_j\rangle$, 
the projection coefficients $C_n = \langle\psi_n|\psi(0)\rangle$ are 
given by
\begin{equation}
\begin{gathered}
C_1 = \frac{C_R + C_L}{\sqrt{2}}, \\
C_2 = N_2\left[\frac{i\sqrt{2}\,\Delta_{a\gamma}}{B}(C_R - C_L) + C_a\right], \\
C_3 = N_3\left[-\frac{i\sqrt{2}\,\Delta_{a\gamma}}{B'}(C_R - C_L) + C_a\right].
\end{gathered}
\label{eq:C123}
\end{equation}
The state of the system after propagating a distance $z$ can be 
expanded as $|\psi(z)\rangle = \sum_{n=1}^{3} C_n\, e^{-iE_n z}|\psi_n
\rangle$. 
We focus on the case with no initial axion component, i.e. $C_a = 0$. 
Substituting the projection coefficients and eigenstates into the 
expansion and rearranging, the amplitudes of the right- and left-handed 
circular polarization components are obtained as
\begin{equation}
\begin{aligned}
    R(z) &= \frac{C_R + C_L}{2}\,e^{-iE_1 z} \\
    &\quad + A\,e^{-iE_2 z}(C_R - C_L) + A'\,e^{-iE_3 z}(C_R - C_L), \\
    L(z) &= \frac{C_R + C_L}{2}\,e^{-iE_1 z} \\
    &\quad - A\,e^{-iE_2 z}(C_R - C_L) - A'\,e^{-iE_3 z}(C_R - C_L),
\end{aligned}
\label{eq:RL}
\end{equation}
where
\begin{equation*}
    A = \frac{2\Delta_{a\gamma}^2}{B^2 + 4\Delta_{a\gamma}^2}, \quad 
    A' = \frac{2\Delta_{a\gamma}^2}{B'^2 + 4\Delta_{a\gamma}^2}.
\end{equation*}
The circular-polarization degree can be written as 
\begin{equation}
\begin{aligned}
    P_C(z
    )&= \frac{|R(z)|^2 - |L(z)|^2}{|R(z)|^2 + |L(z)|^2} \\[4pt]
        &= \frac{2\,\mathrm{Re}(\alpha\eta^*) 
                + 2\,\mathrm{Re}(\alpha\delta^*)}
               {|\alpha|^2 + |\eta|^2 + |\delta|^2 
                + 2\,\mathrm{Re}(\eta\delta^*)},
\end{aligned}
\label{eq:a5}
\end{equation}
where
\begin{equation*}
\begin{gathered}
\alpha = \frac{C_R + C_L}{2}\,e^{-iE_1 z}, \\
\eta = A\,e^{-iE_2 z}(C_R - C_L), \\
\delta = A'\,e^{-iE_3 z}(C_R - C_L).
\end{gathered}
\label{eq:alpha_eta_delta}
\end{equation*}
For an initially linearly polarized photon with polarization angle $\beta$,
\begin{equation}
    C_R = \frac{e^{i\beta}}{\sqrt{2}}, \quad 
    C_L = \frac{e^{-i\beta}}{\sqrt{2}},
\end{equation}
Eq.~\eqref{eq:a5} reduces to the analytic expression given in Eq.~\eqref{12}. For an initially purely circularly polarized photon state, it reduces to Eq.~\eqref{13}.

\section{Analytical Derivation of Circular Polarization Degree in Weak Coupling Approximation via Second-Order Perturbation Theory}
\label{B}
In the axion-photon mixing system,  a weak-mixing regime is established when the detuning greatly exceeds the coupling strength (\(2\Delta_{a\gamma} \ll |\Delta_a - \Delta_{\text{pl}}|\)), allowing for a perturbative treatment.   Starting from the perturbative Hamiltonian, we derive the analytical expression for the circular polarization degree. In the chiral basis $\psi = (|R\rangle, |L\rangle, |a\rangle)^{\mathrm{T}}$, the total Hamiltonian is decomposed into the unperturbed part $H_0$ and the perturbation $H'$:
\begin{equation}
\begin{gathered}
H_0 = \begin{pmatrix}
\Delta_{\mathrm{pl}} & 0 & 0 \\
0 & \Delta_{\mathrm{pl}} & 0 \\
0 & 0 & \Delta_a
\end{pmatrix}, \\[8pt]
H' = \begin{pmatrix}
0 & 0 & \frac{i}{\sqrt{2}}\Delta_{a\gamma} \\
0 & 0 & -\frac{i}{\sqrt{2}}\Delta_{a\gamma} \\
-\frac{i}{\sqrt{2}}\Delta_{a\gamma} & \frac{i}{\sqrt{2}}\Delta_{a\gamma} & 0
\end{pmatrix},
\end{gathered}
\label{eq:hamiltonians}
\end{equation}
where $\Delta_{\mathrm{pl}} = -\omega_{\mathrm{pl}}^2/(2E)$ is the plasma contribution, $\Delta_a = -m_a^2/(2E)$ is the axion mass term, and $\Delta_{a\gamma} = g_{a\gamma\gamma}B_T/2$ is the axion-photon coupling term, following the notation used in Sec.~\ref{sec:Analytic model}. The eigenstates of the unperturbed Hamiltonian $H_0$ are denoted as $|R\rangle$, $|L\rangle$, and $|a\rangle$, with the corresponding zeroth-order eigenvalues $E_1^{(0)} = E_2^{(0)} = \Delta_{\mathrm{pl}}$ and $E_3^{(0)} = \Delta_a$, respectively. Since all diagonal elements of $H'$ vanish, the first-order energy corrections for all states are zero. Thus, we need to compute the second-order energy corrections.

The axion state $|a\rangle$ is non-degenerate  and the standard non-degenerate perturbation formula is
\begin{equation}
E_3^{(2)} = \sum_{n \neq 3} \frac{|\langle n | H' | 3 \rangle|^2}{E_3^{(0)} - E_n^{(0)}}.
\end{equation}
Substituting the non-zero matrix elements $\langle 1|H'|3\rangle = i\Delta_{a\gamma}/\sqrt{2}$ and $\langle 2|H'|3\rangle =- i\Delta_{a\gamma}/\sqrt{2}$, we obtain
\begin{equation}
E_3^{(2)} = \frac{\Delta_{a\gamma}^2}{\Delta_a - \Delta_{\mathrm{pl}}}.
\end{equation}
The photon states $|R\rangle$ and $|L\rangle$ are degenerate under $H_0$, so they cannot be treated with non-degenerate perturbation theory directly. We construct an effective Hamiltonian $H_{\mathrm{eff}}$ within this two-dimensional degenerate subspace to obtain the second-order corrections. The matrix elements of the effective Hamiltonian are given by
\begin{equation}
H_{ij}^{\text{eff}} = \sum_{n \neq 1,2} \frac{\langle i | H' | n \rangle \langle n | H' | j \rangle}{E^{(0)} - E_n^{(0)}}, \qquad i, j \in \{1, 2\},
\label{eq:eff}
\end{equation}
where $E^{(0)}=\Delta_{\mathrm{pl}}$ is doubly degenerate. Taking $|a\rangle$ as the only intermediate state, the effective Hamiltonian simplifies to
\begin{equation}
H_{\mathrm{eff}} = \frac{\Delta_{a\gamma}^2}{2(\Delta_{\mathrm{pl}} - \Delta_a)} \begin{pmatrix}
1 & -1 \\
-1 & 1
\end{pmatrix}.
\end{equation}
Diagonalizing $H_{\mathrm{eff}}$, the second-order energy corrections within the degenerate subspace are
\begin{equation}
E_1^{(2)} = 0, \qquad E_2^{(2)} = \frac{\Delta_{a\gamma}^2}{\Delta_{\mathrm{pl}} - \Delta_a}.
\end{equation}
The corresponding normalized eigenstates are
\begin{equation}
|\psi_1\rangle = \frac{1}{\sqrt{2}} (|R\rangle + |L\rangle), \qquad |\psi_2\rangle = \frac{1}{\sqrt{2}} (|R\rangle - |L\rangle).
\end{equation}
With the eigenvalues and eigenstates of the weakly mixed limit now fully determined, we solve for the system's  evolution. Consider an initial state with no axion component:
\begin{equation}
|\psi_0\rangle = C_R |R\rangle + C_L |L\rangle.
\end{equation}
Expanding the initial state in the eigenstate basis is
\begin{equation}
|\psi_0\rangle = \frac{C_R + C_L}{\sqrt{2}} |\psi_1\rangle + \frac{C_R - C_L}{\sqrt{2}} |\psi_2\rangle.
\end{equation}
Since each eigenmode evolves independently in the eigenbasis, after propagating through distance $z$, the state vector becomes
\begin{equation}
|\psi(z)\rangle = \frac{C_R + C_L}{\sqrt{2}} e^{-iE_1 z} |\psi_1\rangle + \frac{C_R - C_L}{\sqrt{2}} e^{-iE_2 z} |\psi_2\rangle,
\end{equation}
where $E_1 = \Delta_{\mathrm{pl}}$ and $E_2 = \Delta_{\mathrm{pl}} + \dfrac{\Delta_{a\gamma}^2}{\Delta_{\mathrm{pl}} - \Delta_a}$. Substituting the eigenstates $|\psi_1\rangle$ and $|\psi_2\rangle$ and simplifying, the amplitudes of the right- and left-circularly polarization components are obtained as
\begin{align}
R(z) &= \frac{(C_R + C_L)e^{-iE_1 z} + (C_R - C_L)e^{-iE_2 z}}{2}, \\
L(z) &= \frac{(C_R + C_L)e^{-iE_1 z} - (C_R - C_L)e^{-iE_2 z}}{2}.
\end{align}
Applying the definition of circular polarization degree from Eq.~\eqref{11} to the case of purely right-handed circular polarized incidence, i.e. $C_R = 1$, $C_L = 0$, the circular polarization degree simplifies to
\begin{equation}
P_C(z) = \cos\left( \frac{\Delta_{a\gamma}^2}{\Delta_{\mathrm{pl}} - \Delta_a} z \right).
\end{equation}

The above result indicates that, in the weak-mixing limit, the polarization degree of purely circularly polarized incident light undergoes simple harmonic oscillation with a characteristic frequency $\frac{\Delta_{a\gamma}^2}{|\Delta_a - \Delta_{\mathrm{pl}}|}$. This expression is consistent with the leading term of Eq.~(21) in the main text in the limit $\theta \to 0$, providing an independent check of the general analytic solution derived in Sec.~\ref{sec:Analytic model}.

\bibliography{refs}

@article{Finelli:2008jv,
    author = "Finelli, Fabio and Galaverni, Matteo",
    title = "{Rotation of Linear Polarization Plane and Circular Polarization from Cosmological Pseudo-Scalar Fields}",
    eprint = "0802.4210",
    archivePrefix = "arXiv",
    primaryClass = "astro-ph",
    doi = "10.1103/PhysRevD.79.063002",
    journal = "Phys. Rev. D",
    volume = "79",
    pages = "063002",
    year = "2009"
}

@phdthesis{Cadamuro:2012rm,
    author = "Cadamuro, Davide",
    title = "{Cosmological limits on axions and axion-like particles}",
    eprint = "1210.3196",
    archivePrefix = "arXiv",
    primaryClass = "hep-ph",
    reportNumber = "MPP-2012-140",
    school = "Munich U.",
    year = "2012"
}

@article{maiani1986effects,
  title={Effects of nearly massless, spin-zero particles on light propagation in a magnetic field},
  author={Maiani, Luciano and Petronzio, R and Zavattini, E},
  journal={Physics Letters B},
  volume={175},
  number={3},
  pages={359--363},
  year={1986},
  publisher={Elsevier}
}

@article{Bassan:2010ya,
    author = "Bassan, Nicola and Mirizzi, Alessandro and Roncadelli, Marco",
    title = "{Axion-like particle effects on the polarization of cosmic high-energy gamma sources}",
    eprint = "1001.5267",
    archivePrefix = "arXiv",
    primaryClass = "astro-ph.HE",
    doi = "10.1088/1475-7516/2010/05/010",
    journal = "JCAP",
    volume = "05",
    pages = "010",
    year = "2010"
}

@article{du2018search,
  title={Search for invisible axion dark matter with the axion dark matter experiment},
  author={Du, Nick and Force, N and Khatiwada, R and Lentz, E and Ottens, R and Rosenberg, LJ and Rybka, Gray and Carosi, G and Woollett, N and Bowring, D and others},
  journal={Physical review letters},
  volume={120},
  number={15},
  pages={151301},
  year={2018},
  publisher={APS}
}

@article{bahre2013any,
  title={Any light particle search II—technical design report},
  author={B{\"a}hre, Robin and D{\"o}brich, Babette and Dreyling-Eschweiler, Jan and Ghazaryan, Samvel and Hodajerdi, Reza and Horns, Dieter and Januschek, Friederike and Knabbe, E-A and Lindner, Axel and Notz, Dieter and others},
  journal={Journal of Instrumentation},
  volume={8},
  number={09},
  pages={T09001--T09001},
  year={2013}
}

@article{anastassopoulos2017new,
  title={New CAST limit on the axion-photon interaction},
  author={Anastassopoulos, V and Aune, S and Barth, K and Belov, A and Cantatore, GIOVANNI and Carmona, JM and Castel, JF and Cetin, SA and Christensen, F and Collar, JI and others},
  journal={arXiv preprint arXiv:1705.02290},
  year={2017}
}

@article{Reynes:2021bpe,
    author = "Reyn{\'e}s, J{\'u}lia Sisk and Matthews, James H. and Reynolds, Christopher S. and Russell, Helen R. and Smith, Robyn N. and Marsh, M. C. David",
    title = "{New constraints on light axion-like particles using Chandra transmission grating spectroscopy of the powerful cluster-hosted quasar H1821+643}",
    eprint = "2109.03261",
    archivePrefix = "arXiv",
    primaryClass = "astro-ph.HE",
    doi = "10.1093/mnras/stab3464",
    journal = "Mon. Not. Roy. Astron. Soc.",
    volume = "510",
    number = "1",
    pages = "1264--1277",
    year = "2021"
}

@article{Reynolds:2019uqt,
    author = "Reynolds, Christopher S. and Marsh, M. C. David and Russell, Helen R. and Fabian, Andrew C. and Smith, Robyn and Tombesi, Francesco and Veilleux, Sylvain",
    title = "{Astrophysical limits on very light axion-like particles from Chandra grating spectroscopy of NGC 1275}",
    eprint = "1907.05475",
    archivePrefix = "arXiv",
    primaryClass = "hep-ph",
    doi = "10.3847/1538-4357/ab6a0c",
    journal = "Astrophys. J.",
    volume = "890",
    pages = "59",
    year = "2020"
}

@article{Dekker:2025vcg,
    author = "Dekker, Ariane and Herrera, Gonzalo and Kantzas, Dimitrios",
    title = "{Axion-like particle limits from multi-messenger sources}",
    eprint = "2506.14659",
    archivePrefix = "arXiv",
    primaryClass = "hep-ph",
    month = "6",
    year = "2025"
}

@article{Chiba:2025hgr,
    author = "Chiba, Wataru and Jinno, Ryusuke and Nomura, Kimihiro",
    title = "{Axion-photon conversion in stochastic magnetic fields}",
    eprint = "2512.21108",
    archivePrefix = "arXiv",
    primaryClass = "hep-ph",
    reportNumber = "KOBE-COSMO-25-20, KUNS-3086",
    month = "12",
    year = "2025"
}

@article{volvach2016non,
  title={Non-stationary emission of the blazar S4 0954+ 658 over a wide range of wavelength},
  author={Volvach, AE and Bychkova, VS and Larionov, MG and Kardashev, NS and Volvach, LN and Vlasyuk, VV and Spiridonova, OI and L{\"a}hteenm{\"a}ki, A and Tornikoski, Merja and Aller, MF and others},
  journal={Astronomy Reports},
  volume={60},
  number={12},
  pages={1035--1045},
  year={2016},
  publisher={Springer}
}

@article{Jorstad:2017bga,
    author = "Jorstad, Svetlana G. and others",
    title = "{Kinematics of Parsec-Scale Jets of Gamma-Ray Blazars at 43{\textasciitilde}GHz within the VLBA-BU-BLAZAR Program}",
    eprint = "1711.03983",
    archivePrefix = "arXiv",
    primaryClass = "astro-ph.GA",
    doi = "10.3847/1538-4357/aa8407",
    journal = "Astrophys. J.",
    volume = "846",
    pages = "98",
    year = "2017"
}

@article{Roncadelli:2012ar,
    author = "Roncadelli, Marco and De Angelis, Alessandro and Galanti, Giorgio",
    editor = "Oberauer, Lothar and Raffelt, Georg and Wagner, Robert",
    title = "{Importance of axion-like particles for very-high-energy astrophysics}",
    eprint = "1207.0328",
    archivePrefix = "arXiv",
    primaryClass = "astro-ph.HE",
    doi = "10.1088/1742-6596/375/1/052029",
    journal = "J. Phys. Conf. Ser.",
    volume = "375",
    pages = "052029",
    year = "2012"
}

@article{Galanti:2024lfn,
    author = "Galanti, Giorgio",
    title = "{Axion-like Particle Effects on Photon Polarization in High-Energy Astrophysics}",
    eprint = "2407.21421",
    archivePrefix = "arXiv",
    primaryClass = "hep-ph",
    doi = "10.3390/universe10080312",
    journal = "Universe",
    volume = "10",
    number = "8",
    pages = "312",
    year = "2024"
}

@article{Yao:2026yez,
    author = "Yao, Run-Min and Bi, Xiao-Jun and Yin, Peng-Fei and Huang, Qing-Guo",
    title = "{Resonant Photon-Axion Mixing Driven by Dark Matter Oscillations}",
    eprint = "2601.02115",
    archivePrefix = "arXiv",
    primaryClass = "hep-ph",
    month = "1",
    year = "2026"
}

@article{Yao:2022col,
    author = "Yao, Run-Min and Bi, Xiao-Jun and Wang, Jin-Wei and Yin, Peng-Fei",
    title = "{Optical circular polarization induced by axionlike particles in blazars}",
    eprint = "2209.14214",
    archivePrefix = "arXiv",
    primaryClass = "astro-ph.HE",
    doi = "10.1103/PhysRevD.107.043031",
    journal = "Phys. Rev. D",
    volume = "107",
    number = "4",
    pages = "043031",
    year = "2023"
}

@article{Masaki:2017aea,
    author = "Masaki, Emi and Aoki, Arata and Soda, Jiro",
    title = "{Photon-Axion Conversion, Magnetic Field Configuration, and Polarization of Photons}",
    eprint = "1702.08843",
    archivePrefix = "arXiv",
    primaryClass = "astro-ph.CO",
    reportNumber = "KOBE-COSMO-17-02",
    doi = "10.1103/PhysRevD.96.043519",
    journal = "Phys. Rev. D",
    volume = "96",
    number = "4",
    pages = "043519",
    year = "2017"
}

@article{Ghisellini:2009fj,
    author = "Ghisellini, G. and Tavecchio, F. and Foschini, L. and Ghirlanda, G. and Maraschi, L. and Celotti, A.",
    title = "{General physical properties of bright Fermi blazars}",
    eprint = "0909.0932",
    archivePrefix = "arXiv",
    primaryClass = "astro-ph.CO",
    doi = "10.1111/j.1365-2966.2009.15898.x",
    journal = "Mon. Not. Roy. Astron. Soc.",
    volume = "402",
    pages = "497",
    year = "2010"
}

@article{Wang:2025hyu,
    author = "Wang, Xianqi and others",
    title = "{Digital electronics for the eXTP large area detector}",
    doi = "10.1007/s10686-025-10017-9",
    journal = "Exper. Astron.",
    volume = "60",
    number = "1",
    pages = "5",
    year = "2025"
}

@article{ratheesh2021imaging,
  title={The Imaging X-Ray Polarimetry Explorer (IXPE): Pre-Launch},
  author={Ratheesh, Ajay and Rubini, Alda and Marscher, Alan and Manfreda, Alberto and Marrocchesi, Alessandra and Brez, Alessandro and Di Marco, Alessandro and Paggi, Alessandro and Profeti, Alessandro and Nuti, Alessio and others},
  journal={arXiv e-prints},
  pages={arXiv--2112},
  year={2021}
}

@article{Soffitta:2013hla,
    author = "Soffitta, Paolo and others",
    title = "{XIPE: the X-ray Imaging Polarimetry Explorer}",
    eprint = "1309.6995",
    archivePrefix = "arXiv",
    primaryClass = "astro-ph.HE",
    doi = "10.1007/s10686-013-9344-3",
    journal = "Exper. Astron.",
    volume = "36",
    pages = "523--567",
    year = "2013"
}

@article{Kole:2024ekt,
    author = "Kole, Merlin and others",
    title = "{Response of the first POLAR-2 prototype to polarized beams}",
    eprint = "2406.05783",
    archivePrefix = "arXiv",
    primaryClass = "astro-ph.IM",
    doi = "10.1088/1748-0221/19/08/P08002",
    journal = "JINST",
    volume = "19",
    number = "08",
    pages = "P08002",
    year = "2024"
}

@article{Bhalerao:2016lex,
    author = "Bhalerao, V. and others",
    title = "{The Cadmium Zinc Telluride Imager on AstroSat}",
    eprint = "1608.03408",
    archivePrefix = "arXiv",
    primaryClass = "astro-ph.IM",
    doi = "10.1007/s12036-017-9447-8",
    journal = "J. Astrophys. Astron.",
    volume = "38",
    pages = "31",
    year = "2017"
}

@article{Lopez-Oramas:2025vld,
    author = "L{\'o}pez-Oramas, Alicia",
    collaboration = "CTAO",
    title = "{CTAO status and perspective}",
    doi = "10.1051/epjconf/202531901002",
    journal = "EPJ Web Conf.",
    volume = "319",
    pages = "01002",
    year = "2025"
}

@article{Mazin:2019ykz,
    author = "Mazin, Daniel",
    collaboration = "CTA Consortium",
    title = "{The Cherenkov Telescope Array}",
    eprint = "1907.08530",
    archivePrefix = "arXiv",
    primaryClass = "astro-ph.IM",
    doi = "10.22323/1.358.0741",
    journal = "PoS",
    volume = "ICRC2019",
    pages = "741",
    year = "2020"
}

@article{Clausen-Brown:2011upb,
    author = "Clausen-Brown, Eric and Lyutikov, Maxim and Kharb, Preeti",
    title = "{Signatures of large-scale magnetic fields in AGN jets: transverse asymmetries}",
    eprint = "1101.5149",
    archivePrefix = "arXiv",
    primaryClass = "astro-ph.HE",
    doi = "10.1111/j.1365-2966.2011.18757.x",
    journal = "Mon. Not. Roy. Astron. Soc.",
    volume = "415",
    pages = "2081",
    year = "2011"
}

@article{Meyer:2014epa,
    author = "Meyer, Manuel and Montanino, Daniele and Conrad, Jan",
    title = "{On detecting oscillations of gamma rays into axion-like particles in turbulent and coherent magnetic fields}",
    eprint = "1406.5972",
    archivePrefix = "arXiv",
    primaryClass = "astro-ph.HE",
    doi = "10.1088/1475-7516/2014/09/003",
    journal = "JCAP",
    volume = "09",
    pages = "003",
    year = "2014"
}

@article{Tavecchio:2014yoa,
    author = "Tavecchio, Fabrizio and Roncadelli, Marco and Galanti, Giorgio",
    title = "{Photons to axion-like particles conversion in Active Galactic Nuclei}",
    eprint = "1406.2303",
    archivePrefix = "arXiv",
    primaryClass = "astro-ph.HE",
    doi = "10.1016/j.physletb.2015.04.017",
    journal = "Phys. Lett. B",
    volume = "744",
    pages = "375--379",
    year = "2015"
}

@article{Meyer:2021pbp,
    author = "Meyer, Manuel and Davies, James and Kuhlmann, Julian",
    title = "{gammaALPs: An open-source python package for computing photon-axion-like-particle oscillations in astrophysical environments}",
    eprint = "2108.02061",
    archivePrefix = "arXiv",
    primaryClass = "astro-ph.HE",
    doi = "10.22323/1.395.0557",
    journal = "PoS",
    volume = "ICRC2021",
    pages = "557",
    year = "2021"
}

@article{Liodakis:2023qnt,
    author = "Liodakis, I. and others",
    title = "{Optical circular polarization of blazar S4 0954+65 during high linear polarized states}",
    eprint = "2311.03450",
    archivePrefix = "arXiv",
    primaryClass = "astro-ph.HE",
    doi = "10.1051/0004-6361/202348214",
    journal = "Astron. Astrophys.",
    volume = "680",
    pages = "L11",
    year = "2023"
}

@article{Jansson:2012pc,
    author = "Jansson, Ronnie and Farrar, Glennys R.",
    title = "{A New Model of the Galactic Magnetic Field}",
    eprint = "1204.3662",
    archivePrefix = "arXiv",
    primaryClass = "astro-ph.GA",
    doi = "10.1088/0004-637X/757/1/14",
    journal = "Astrophys. J.",
    volume = "757",
    pages = "14",
    year = "2012"
}

@article{cordes2002ne2001,
  title={NE2001. I. A new model for the galactic distribution of free electrons and its fluctuations},
  author={Cordes, James M and Lazio, T Joseph W},
  journal={arXiv preprint astro-ph/0207156},
  year={2002}
}

@article{Zhu:2024kmu,
    author = "Zhu, Ben-Yang and Huang, Xiaoyuan and Yin, Peng-Fei",
    title = "{Constraints on axion-like particles from the gamma-ray observation of the Galactic Center}",
    eprint = "2408.12234",
    archivePrefix = "arXiv",
    primaryClass = "astro-ph.HE",
    doi = "10.1088/1475-7516/2025/01/030",
    journal = "JCAP",
    volume = "01",
    pages = "030",
    year = "2025"
}

@article{Malyshev:2025iis,
    author = "Malyshev, Denys and Zadorozhna, Lidiia and Bidasyuk, Yuriy and Santangelo, Andrea and Ruchayskiy, Oleg",
    title = "{Constraints on axion-like particles from active galactic nuclei seen through galaxy clusters}",
    eprint = "2506.02848",
    archivePrefix = "arXiv",
    primaryClass = "astro-ph.HE",
    doi = "10.1038/s41550-025-02621-8",
    journal = "Nature Astron.",
    volume = "9",
    number = "9",
    pages = "1387--1395",
    year = "2025"
}

@article{Zhang:2024jvq,
    author = "Zhang, Qingxiang and Huang, Feng and Wang, Zhongxiang and Fang, Taotao",
    title = "{X-ray polarimetric features of gamma-ray bursts across varied redshifts and hints for axionlike particles}",
    eprint = "2404.07555",
    archivePrefix = "arXiv",
    primaryClass = "astro-ph.HE",
    doi = "10.1103/PhysRevD.111.023028",
    journal = "Phys. Rev. D",
    volume = "111",
    number = "2",
    pages = "023028",
    year = "2025"
}

@article{Benabou:2025jcv,
    author = "Benabou, Joshua N. and Dessert, Christopher and Patra, Kishore C. and Brink, Thomas G. and Zheng, WeiKang and Filippenko, Alexei V. and Safdi, Benjamin R.",
    title = "{Search for Axions in Magnetic White Dwarf Polarization at Lick and Keck Observatories}",
    eprint = "2504.12377",
    archivePrefix = "arXiv",
    primaryClass = "hep-ph",
    month = "4",
    year = "2025"
}

@article{Song:2024rru,
    author = "Song, Ningqiang and Su, Liangliang and Wu, Lei",
    title = "{Polarization signals from axion-photon resonant conversion in a neutron star magnetosphere}",
    eprint = "2402.15144",
    archivePrefix = "arXiv",
    primaryClass = "hep-ph",
    doi = "10.1103/PhysRevD.111.043025",
    journal = "Phys. Rev. D",
    volume = "111",
    number = "4",
    pages = "043025",
    year = "2025"
}

@article{betancourt2025polarization,
  title={Polarization measurements as a probe of axion-photon coupling: A study of GRB 221009A},
  author={Betancourt Kamenetskaia, Boris and Fraija, Nissim and Herrera, Gonzalo},
  journal={Physical Review D},
  volume={111},
  number={11},
  pages={115008},
  year={2025},
  publisher={APS}
}

@article{Liu:2019brz,
    author = "Liu, Tao and Smoot, George and Zhao, Yue",
    title = "{Detecting axionlike dark matter with linearly polarized pulsar light}",
    eprint = "1901.10981",
    archivePrefix = "arXiv",
    primaryClass = "astro-ph.CO",
    doi = "10.1103/PhysRevD.101.063012",
    journal = "Phys. Rev. D",
    volume = "101",
    number = "6",
    pages = "063012",
    year = "2020"
}

@article{liu2023pulsar,
  title={Pulsar polarization arrays},
  author={Liu, Tao and Lou, Xuzixiang and Ren, Jing},
  journal={Physical Review Letters},
  volume={130},
  number={12},
  pages={121401},
  year={2023},
  publisher={APS}
}

@article{PPTA:2024mgh,
    author = "Xue, Xiao and others",
    collaboration = "PPTA",
    title = "{Pulsar Polarization Array Limits on Ultralight Axionlike Dark Matter}",
    eprint = "2412.02229",
    archivePrefix = "arXiv",
    primaryClass = "astro-ph.HE",
    doi = "10.1103/mptv-3x6g",
    journal = "Phys. Rev. Lett.",
    volume = "136",
    number = "1",
    pages = "011001",
    year = "2026"
}

@article{li2026probing,
  title={Probing ultralight axionlike dark matter: A pulsar timing arrays-pulsar polarization arrays synergy},
  author={Li, Ximeng and Liu, Yonghao and Chen, Zu-Cheng and Dai, Shi and Goncharov, Boris and Hu, Xiao-Song and Huang, Qing-Guo and Liu, Tao and Ren, Jing and Wu, Yu-Mei and others},
  journal={Physical Review D},
  volume={113},
  number={4},
  pages={043059},
  year={2026},
  publisher={APS}
}

@article{Wang:2024sdz,
    author = "Wang, Bao and Yang, Xuan and Wei, Jun-Jie and Zhang, Song-Bo and Wu, Xue-Feng",
    title = "{Detecting extragalactic axion-like dark matter with polarization measurements of fast radio bursts}",
    eprint = "2402.00473",
    archivePrefix = "arXiv",
    primaryClass = "astro-ph.HE",
    doi = "10.1038/s42005-025-02045-w",
    journal = "Commun. Phys.",
    volume = "8",
    number = "1",
    pages = "130",
    year = "2025"
}

@article{Huang:2025rrb,
    author = "Huang, Qiu-Ju and Wang, Bao and Wei, Jun-Jie and Wu, Xue-Feng",
    title = "{Hunting for Extragalactic Axion-like Dark Matter in a Decade-long Blazar Optical Polarimetry}",
    eprint = "2511.05839",
    archivePrefix = "arXiv",
    primaryClass = "astro-ph.HE",
    month = "11",
    year = "2025"
}

@article{POLARBEAR:2025djl,
    author = "Adkins, Tylor and others",
    collaboration = "POLARBEAR",
    title = "{Constraints on the polarization angle oscillations of the Crab Nebula with the Simons Array and its applications to the search for axionlike particles}",
    eprint = "2512.18882",
    archivePrefix = "arXiv",
    primaryClass = "astro-ph.CO",
    doi = "10.1103/9wqc-tgzq",
    journal = "Phys. Rev. D",
    volume = "113",
    number = "4",
    pages = "043044",
    year = "2026"
}

@article{PhysRevD.37.1237,
  title = {Mixing of the photon with low-mass particles},
  author = {Raffelt, Georg and Stodolsky, Leo},
  journal = {Phys. Rev. D},
  volume = {37},
  issue = {5},
  pages = {1237--1249},
  numpages = {0},
  year = {1988},
  month = {Mar},
  publisher = {American Physical Society},
  doi = {10.1103/PhysRevD.37.1237},
  url = {https://link.aps.org/doi/10.1103/PhysRevD.37.1237}
}

@article{Dessert:2022yqq,
    author = "Dessert, Christopher and Dunsky, David and Safdi, Benjamin R.",
    title = "{Upper limit on the axion-photon coupling from magnetic white dwarf polarization}",
    eprint = "2203.04319",
    archivePrefix = "arXiv",
    primaryClass = "hep-ph",
    doi = "10.1103/PhysRevD.105.103034",
    journal = "Phys. Rev. D",
    volume = "105",
    number = "10",
    pages = "103034",
    year = "2022"
}

@article{Noordhuis:2022ljw,
    author = "Noordhuis, Dion and Prabhu, Anirudh and Witte, Samuel J. and Chen, Alexander Y. and Cruz, F\'abio and Weniger, Christoph",
    title = "{Novel Constraints on Axions Produced in Pulsar Polar-Cap Cascades}",
    eprint = "2209.09917",
    archivePrefix = "arXiv",
    primaryClass = "hep-ph",
    doi = "10.1103/PhysRevLett.131.111004",
    journal = "Phys. Rev. Lett.",
    volume = "131",
    number = "11",
    pages = "111004",
    year = "2023"
}

@article{Peccei:1977hh,
    author = "Peccei, R. D. and Quinn, Helen R.",
    title = "{CP Conservation in the Presence of Instantons}",
    reportNumber = "ITP-568-STANFORD",
    doi = "10.1103/PhysRevLett.38.1440",
    journal = "Phys. Rev. Lett.",
    volume = "38",
    pages = "1440--1443",
    year = "1977"
}

@article{Svrcek:2006yi,
    author = "Svrcek, Peter and Witten, Edward",
    title = "{Axions In String Theory}",
    eprint = "hep-th/0605206",
    archivePrefix = "arXiv",
    reportNumber = "SLAC-PUB-11894",
    doi = "10.1088/1126-6708/2006/06/051",
    journal = "JHEP",
    volume = "06",
    pages = "051",
    year = "2006"
}

@article{Dessert:2020lil,
    author = "Dessert, Christopher and Foster, Joshua W. and Safdi, Benjamin R.",
    title = "{X-ray Searches for Axions from Super Star Clusters}",
    eprint = "2008.03305",
    archivePrefix = "arXiv",
    primaryClass = "hep-ph",
    doi = "10.1103/PhysRevLett.125.261102",
    journal = "Phys. Rev. Lett.",
    volume = "125",
    number = "26",
    pages = "261102",
    year = "2020"
}

\end{document}